\documentclass{article}

\usepackage[letterpaper,top=2cm,bottom=2cm,left=3cm,right=3cm,marginparwidth=1.75cm]{geometry}

\usepackage{amsmath}
\usepackage{graphicx}
\graphicspath{{figureY/}} 
\usepackage{hyperref}
\hypersetup{colorlinks,allcolors=black}
\title{Eigen mode selection in human subject game experiment}
\author{Wang Zhijian and Yao Qinmei and Wang Yijia \\ Experimental social science laboratory, Zhejiang University, China}

\begin{document}

%
%
%
%
%

\maketitle
\begin{abstract}
Eigen mode selection ought to be a practical issue in some real game systems, as it is a practical issue in the dynamics behaviour of a building, bridge, or molecular, because of the mathematical similarity in theory. However, its reality and accuracy have not been known in real games. We design a 5-strategy game which, in the replicator dynamics theory, is predicted to exist two eigen modes. Further, in behaviour game theory, the game is predicted that the mode selection should depends on the game parameter. We conduct human subject game experiments by controlling the parameter. The data confirm that, the predictions on the mode existence, as well as the mode selection are significantly supported. This finding suggests that, like the equilibrium selection concept in classical game theory, eigen mode selection is an issue in game dynamics theory. \end{abstract}
 
\tableofcontents

\newpage

\section{Introduction}

\subsection{Eigenmode}
We investigate the reality and accuracy of the concept, eigen mode, by theory and experiment in this study. Here, the the theory indicates the game dynamics theory (or evolutionary game theory)\cite{2011Sandholm, dan2016} which is a dynamics version in game theory \cite{samuelson2016}; The experiment indicates the laboratory human subjects behaviour game experiments, which is called also as experimental economics approach \cite{Behavioral2003, dan2016}. 

Eigen mode is a normal mode in an oscillating system, being one in which all parts of the system are oscillating with the same frequency and  with a fixed phase relation.  A physical object, such as a building, bridge, or molecule, has a set of normal modes and their natural frequencies that depend on its structure, materials and boundary conditions. In music, normal modes of vibrating instruments (strings, air pipes, drums, etc.) are called "harmonics" or "overtones". The most general motion of a system is a superposition of its normal modes. The modes are normal in the sense that they can move independently, that is to say that an excitation of one mode will never cause motion of a different mode. In mathematical terms, normal modes are orthogonal to each other. Further, in mathematical narrative, eigen mode, like the terms eigenvalue, eigenvector, eigen space, invariant manifold, and so on, belongs to the concept set 'eigen system' in dynamics system theory. 

Game dynamics theory (evolutionary game theory) mathematically belongs to dynamics system theory. Game experimenter have applied  
the eigenvalue to design experiment. E.g., to predict the stability of a game \cite{dan2010tasp, 2015nowak} by the real part of the eigenvalue, and to interpret the cycle by the imaginary part of the eigenvalue \cite{dan2014, wang2014, wang2014social}. 
The logic chain for eigenvalue judgement is following
\begin{itemize}
\item For a give game, the game dynamics system can be expressed as a velocity vector field \cite{2011Sandholm}\cite{dan2016} as 
\begin{equation}
  \dot{x} = f(x) 
\end{equation}
in which $x\in R^N$ and $N$ is the dimension of the strategy space. The Nash equilibrium is a singular point of the velocity vector field (a point
where $\dot{x} = 0$). In linear approximation, near the equilibrium point, the dynamics can be expressed as
$  \dot{x} = J x$ 
in which $J$ is the Jacobian (character matrix) at the equilibrium point \cite{2011Sandholm, dan2016}. An eigenvalue ($\lambda_i$) is a root  of the characteristic equation $(J - \lambda_i I = 0)$. Associated to the eigenvalue, there exists an eigenvector $\xi_i$ which describes the instinct motion shape. These are the basic of the concept set of eigen system. 
Suppose that, an initial condition can be expressed as 
  $x(0) = \sum_{i=1}^N a_i {\xi}_i$
then, the evolution trajectory can be expressed as 
\begin{equation}\label{eq:eigdeco2}
x(t) = \sum_{i=1}^N e^{\lambda_1t} a_i {\xi}_i.  
\end{equation}
Here, each of the eigenvector $\xi_i$ describes an eigen mode, which has $N$ components being oscillating with the same frequency $\lambda_i$. In motion shape, the eigenvector $\xi_i$ structure play the cruel role. The set $\{\xi_i\}$ is called eigen mode set.
\item Eigen mode selection should be a practical issue in real game system in theory. Because, a game dynamics system, in mathematics like a physical object, can have an eigen modes set; At the same time, the game dynamics system can be expected as a sum of its eigen modes\footnote{This issue has appeared in quantum physics. In the quantum physics for the electrons of a hydro atom, by Schr\"{o}dinger equation, the eigen function (eigen mode) set can be obtained. But for a given real world condition, how the electron distributes on the eigen mode is remain an issue. To answer this question, the knowledge from statistical physics, like Boltzmann constant, is needed. When such question is answered well, we can, for example, report the temperature of a star far away.}. 
In a given long run high stochastic dynamics process, statistically, which of the eigen mode mostly contribute to an observation, 
this is the question about the eigen mode selection. 
\end{itemize} 

\subsection{Research question}
Game theorists has rare handled the complex eigenvector, let alone eigen mode concept or the multi eigen mode issue. Although recent experiments \cite{2021Qinmei, 2021Shujie} have significantly supported the eigen mode predictions from the logic chain above, but these game have only unique pair of complex eigenvector. Multi eigen mode was seemly test out in significant 2 roles and 4 strategy zero sum game, of which having 3 pairs of complex eigenvectors (eigen modes), in the existed data (\cite{ONeill1987, Binmore2001Minimax, Yoshitaka2013Minimax}) reported by \cite{WY2020}. But the data does not provide control variable for eigenvectors, then the cause for multi eigen mode can not be distinguished. 

Naturally, a naive question will be: When a game have multi eigen modes, which mode will be selected? This question likes that question in game statics theory --- when a system have multi equilibrium, which equilibrium will be selected --- which is the issue namely equilibrium selection. Further question is why a mode been selected? 

We design an ideal controllable 5 strategy game experiment, of which having two eigen modes. By this exemplified game, we show the reality and accuracy of the multi eigen mode selection prediction from theory, and then carry out the verification by real human subject game experiments. 

\section{Results}

\subsection{Theoretical expectation}
\subsubsection{Game and its eigensystem}
 
\paragraph{Payoff matrix}  The payoff matrix of 
the one population symmetric game is   
\begin{equation}\label{eq:payoffmatrix}
  A = 
 \left(\begin{array}{rrrrr}
 0 & a & 1 & -1 & -a\\
 -a & 0 & a & 1 & -1\\
 -1 & -a & 0 & a & 1\\
 1 & -1 & -a & 0 & a\\
 a & 1 & -1 & -a & 0 
 \end{array}\right)
\end{equation} 
\paragraph{Dynamics} 
To investigate the dynamic behaviors in laboratory experiment game,
we begin with using the replicator dynamics equations \cite{taylor1978evolutionary}:
\begin{equation}\label{eq:repliequl}
  \Dot{x}_i= x_i \Big(U_i - {\overline{U}}\Big) 
\end{equation}
in which, $i\in \{1,2,..,N\}$.
$x_i$ is the $i$-th strategy player's probability
in the population where the $i$-th strategy player included, and 
$\Dot{x}_i$ is the evolution velocity of the probability; 
 $U_i$ ($:\equiv \sum_j A_{ij} x_j$) is the payoff of the players using the $i$-th strategy, 
and $\overline{U}$ (:$\equiv x_i U_i$) is the average payoff of
the full population. 
\paragraph{Equilibrium} 
Nash equilibrium $(1/5,1/5,1/5,1/5,1/5)^T$.
\paragraph{Jacobian} This is a time invariant dynamics system. 
Suppose that, the motion of the strategy vector $x$ is close the equilibrium and the linear approximation of 
dynamical system is validate. Then we can obtain the eigen system from the Jacobian, which collects all first-order partial derivatives of the multivariate function (Eq. \ref{eq:repliequl}) at the equilibrium as a character matrix \cite{2011Sandholm, dan2016}: 
\begin{equation}\label{eq:J}
   J = \frac{1}{5} \left(\begin{array}{rrrrr} 0 & a & 1 & -1 & -a\\ -a & 0 & a & 1 & -1\\ -1 & -a & 0 & a & 1\\ 1 & -1 & -a & 0 & a\\ a & 1 & -1 & -a & 0 \end{array}\right)
\end{equation}

\begin{figure}[!ht]
\centering
~~~~~~~~~~~~\includegraphics[scale=0.60]{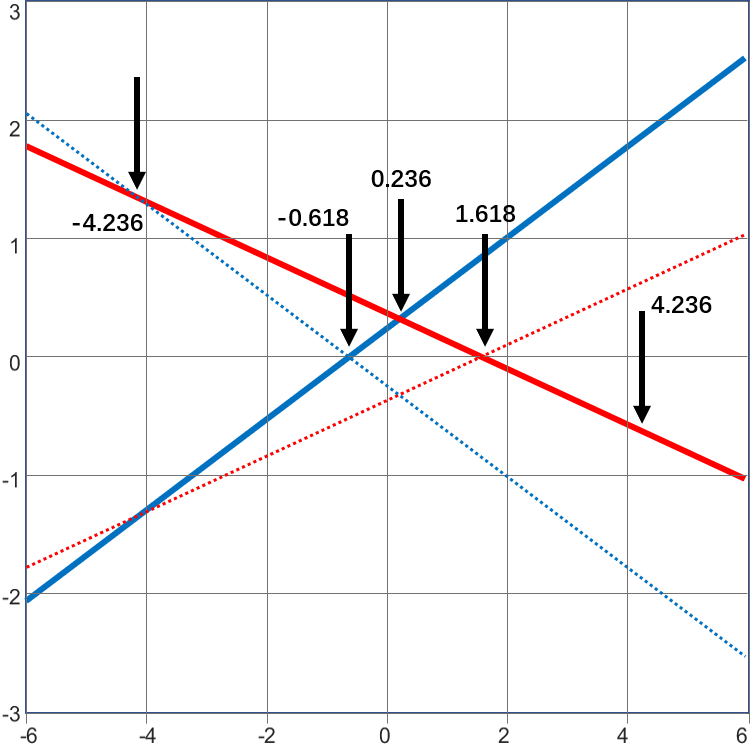} \\
~~~~ \includegraphics[scale=0.60]{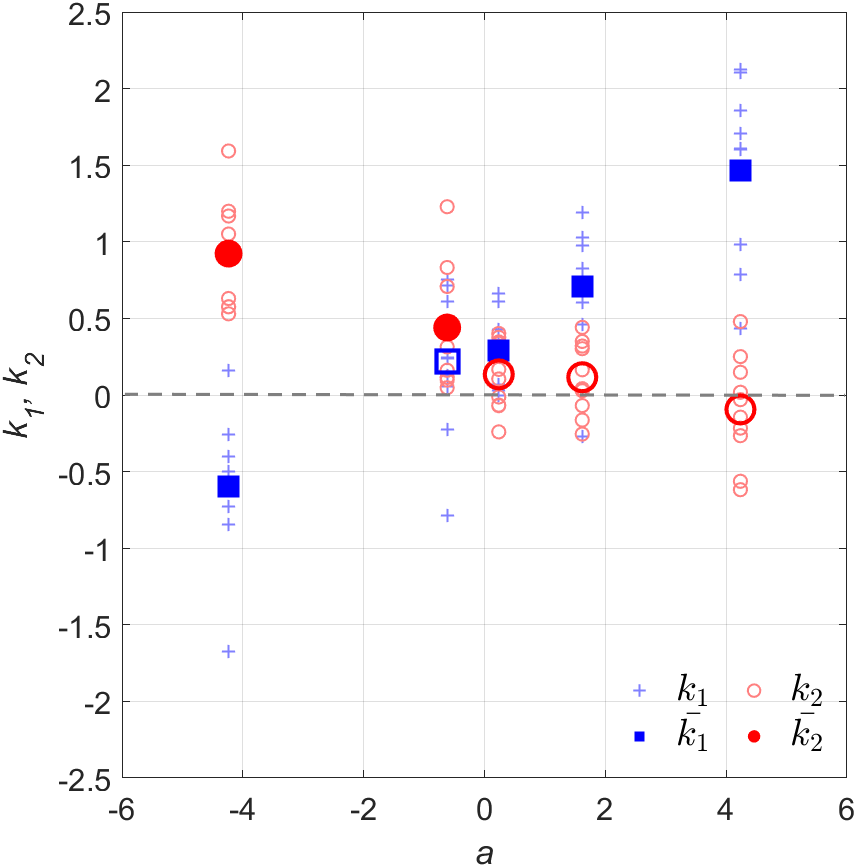} 
\caption{The eigenvalues and observed mode selection. 
Top panel, the imaginary part of the eigenvalues as a function of $a$ in the game matrix. The heavy line indicates the observable eigen mode, referring to the myopic response estimation. The blue (red) indicates 
$\sigma_\alpha$ and $\sigma_\beta$ associated eigenvalues (imaginary part), respectively. The arrows indicates the parameter $a$ values used in experiments. 
Bottom panel, the experimental weight of the mode selection. Results of the session shown in small. A Big label indicates the average of the sessions. Red (and Blue) indicates the $k_1$ (and $k_2$) value. A big label with fulled (not fulled) surface indicates the value is deviated from 0 in significant (not significant) by \textbf{ttest} over the observation from 10 sessions. Numerical results see Table \ref{tab:k1andk2session} in appendix.}
\label{fig:eva_a}
\end{figure}

\paragraph{Eigenvalue} 
The eigenvalues are 
\begin{equation}\label{eq:eigenvalue}
    \lambda = (0, \chi^{+},~ \chi^{-},~ -\chi^{+},~ -\chi^{-}),
\end{equation} 
in which $\chi^{\pm} = \frac{1}{10}\Big[\pm 2\, \sqrt{5} ( 1 - 4\, a - a^2 ) - 10 (a^2 + 1)\Big]^{1/2} $. Notice that, $\chi^{\pm}$ is full imaginary, and their real parts are the constant 0 ($Re(\chi^{\pm}) = 0).$ 
The two pairs of conjunction complex eigenvalue depend on the variable $a$ which can be shown in Fig. \ref{fig:eva_a}. 
The imaginary part of eigenvalue $\Im(\lambda_i)$ relates to 
the angular velocity, 
the frequency, and the angular momentum of cycle motion \cite{wang2014}. It is natural to assume that, when given $a$, a motion  having higher $|\Im(\lambda_i)|$ value has larger angular momentum, if other factors are same.  Moreover, if $\Im(\lambda_i) = 0$, the angular momentum will be 0. Details of relationship of $\Im(\lambda)$ and angular momentum can refer to  the section 6.2.1 Interfere of two components from the same eigenvector in \cite{WY2020}.

\begin{figure}[h!]
\centering
\includegraphics[scale=1.2]{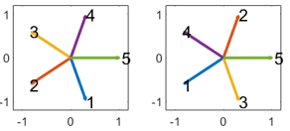}
  \caption{Geometric presentation of the eigenvector (left) $v^\alpha$ and (right) $v^\beta$.}
\label{fig:eig_vec}
\end{figure}

\paragraph{Eigenvector}  There are five eigenvector associated to the five eigenvalues respectively.  
\begin{itemize}
\item Among the five eigenvectors, There is an the unique pure real eigenvector, which is $N_e=(1,1,1,1,1)^T$, the rest point (Nash equilibrium). 
\item Remained are the two conjunction pairs of complex vectors. 
The amplitudes of each eigenvector components are all equal. 
The phase of each eigenvector component are fixed as 
$2\pi/5$ and $4\pi/5$ along strategy-(1,2,3,4,5), respectively. 
The pattern of the two eigenvectors in complex plane is 
shown in Fig. \ref{fig:eig_vec}. 
\item The eigenvector set is orthogonal. 
\end{itemize}

\begin{figure}[h!]
\centering
\includegraphics[scale=0.3]{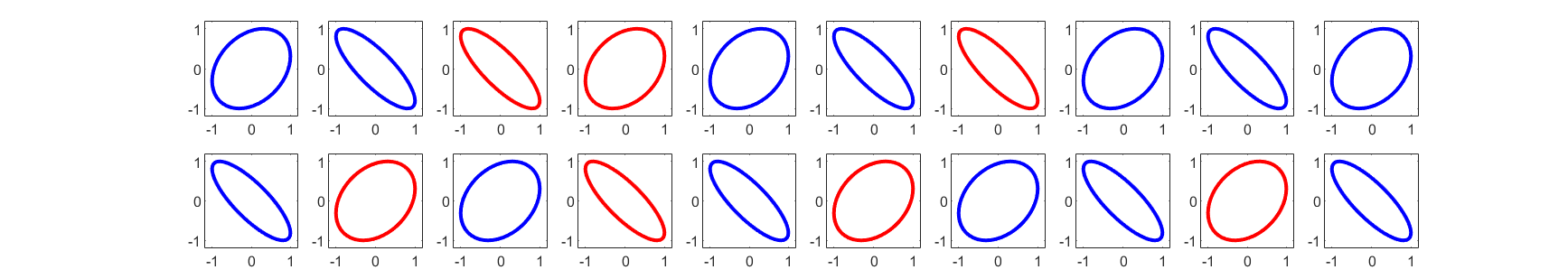}
  \caption{Geometric presentation of the two eigencycle set ((top) $\sigma_\alpha$ and (bottom) $\sigma_\beta$). These two eigencycle set associate to the two eigenvectors (shown in Fig. \ref{fig:eig_vec}) respectively. The orbit is the Lissajous curve in mathematics. Each sub figure represents a 2-d subspace, whose dimension id is arranged as the first column in Table \ref{tab:ec_qualitative}, from left to right ordered as $mn$=(12, 13, 14, 15, 23, 24, 25, 34, 35, 45). The red (or blue) indicates the eigencycle value is positive and counter-clockwise (or negative, clockwise). The eigen mode decomposition refers to these two eigencycle set ($(\sigma_\alpha, \sigma_\beta$). Meanwhile, the equivalent among [1,4,5,8,10]-columns, as well as the equivalent among [2,3,6,7,9]-columns, hint the invariant between the 2-d sub-spaces. \label{fig:eig_lissa}} 
\end{figure}

\paragraph{Eigencycle} An eigencycle is determined by an eigenvector 
(definition is shown in appendix \ref{app:def_ec}). 
The eigencycle of the 10 subspace can be calculated. The results of eigencycle set from the two eigenvectors ($v^\alpha$ and $v^\beta$)
are shown in 
Table \ref{tab:ec_qualitative} as $\sigma_\alpha$ and $\sigma_\beta$. 
These two eigencycle sets, 
reflecting the two eigenvector then the eigen mode. 
The geometric presentation of the two eigencycle set is shown in
Fig. \ref{fig:eig_lissa}. Regarded as two vectors with 10 components, these two vector is orthogonal, because the inner product (dot product) of the two vectors is zero. 
 So called eigen mode selection is the projection of cyclic motion to these two eigencycle set vectors.
 These two eigencycle set
play cruel role of this study.

\subsubsection{Parameter specification and theoretical prediction \label{sec:theo_pred}}
The parameter specification, as well as the related theoretical predictions, plays the cruel role in this experiment study. 
As the first experiment, to our knowledge, to study the eigen mode selection, 
we limit ourselves on following three research questions: 
\begin{enumerate}
\item (RQ1: Distinguishable) Whether the eigen modes exists and distinguishable?
\item (RQ2: Predictable) Whether the eigen modes selection can be predictable?
\item (RQ3: Invariant) Whether the parameter selection invariant exists?
\end{enumerate} 

\paragraph{Parameter specification}
In mathematics, the parameter selection mainly base on the solution of the eigen system, especially on the eigenvalues relationship of the two eigen mode. 
This is because, eigen mode (eigenvector) is the companion of its related eigenvalue in dynamics. If the eigenvalue of an eigenvector close to 0, 
the related cyclic frequency tends to 0; Then, the cyclic motion referring to angular momentum measurement is statistically weak and being no observable, and is regarded as dominated and no selected. This study is based on angular momentum measurement. So, the mode selection is the contribution weight of an eigen mode to total observed angular momentum. In this view, special relationship of the eigenvalues of the two mode, like equal or zero, is cruel. This is the main reasons to specify the parameter $a$.  

In total, we choose the five parameters as $a= [-4.236, -0.618, 0.234, 1.618, 4.236]$, and call the related game treatment in this study as Treatment-(1, 2, 3, 4, 5) respectively, which denoted as (Tr1, Tr2, Tr3, Tr4, Tr5) in short. The reasons of the parameter assignments are to reply the research questions: 
 
\paragraph{Prediction 1}, which is aim at answering the RQ1, we assign the parameter $a=(-0.618, 1.618)$ are chose because at these two value, $\chi^+(-0.618)=0$ and $\chi^-(1.618)=0$; So, only one mode can be observed in these two treatments, respectively. This is the impact of the special condition of 0 eigenvalue on angular momentum measurement.
\paragraph{Prediction 2}, which is aim at the RQ2, we assign the parameter $a=(-4.236, 0.236)$. Because $\big|\chi^+(-4.236)\big|=\big|\chi^-(0.236)\big|$ at these two value, So, the two mode observed should be selected equally in these two treatments, respectively. This is the impact of the special condition of equal eigenvalue on angular momentum measurement.  Meanwhile, for symmetry consideration, 
we choose the $a$ which oppose of the $a=-4.236$. At this point $\big|\chi^+(4.236)\big| >> \big|\chi^-(4.236)\big|$ which is appreciate for the qualitative analysis.

Details of the deduction is shown in appendix secction \ref{app:a_select}.

\begin{table}[!ht] 
 \centering
 \caption{Weight in $mn$-subspace (top), the results of the eigen mode selection of the theory (middle) and 
 the experiment (bottom) 
 \label{tab:ec_qualitative}}
 \begin{tabular}{crrrrrrrrrrr}
\hline 
mn&  $\sigma_a$ & $\sigma_\beta$  & $L$(-4.236)  & $L$(-0.618)& $L$(0.234)& $L$(1.618) & $L$(4.236)\\ 
\hline 
12& -2.988& 1.847& 2& 1/2 & -1/2 & -2 & -2\\
13& -1.847& -2.988& -1&-1 &-1 &-1 & -1\\
14& 1.847& 2.988& 1& 1& 1& 1& 1\\
15& 2.988& -1.847& -2& -1/2& 1/2& 2& 2\\
23& -2.988& 1.847& 2& 1/2& -1/2& -2& -2\\
24& -1.847& -2.988& -1& -1& -1& -1& -1\\
25& 1.847& 2.988& 1& 1& 1& 1& 1\\
34& -2.988& 1.847& 2& 1/2& -1/2& -2& -2\\
35& -1.847& -2.988& -1& -1& -1& -1& -1\\
45& -2.988& 1.847& 2& 1/2& -1/2& -2& -2\\
\hline
~~~~ Theory ($\rho$) &\\  
$\sigma_\alpha$ &  1 & 0.0500 & -0.4543 & 0.1307 & \colorbox[rgb]{1,0.9,0.9}{0.8348}& \colorbox[rgb]{1,0.9,0.9}{0.9950} & \colorbox[rgb]{1,0.9,0.9}{0.9950} \\
 $\sigma_\beta$ &  0.0500 & 1 & \colorbox[rgb]{1,0.9,0.9}{0.8670} & \colorbox[rgb]{1,0.9,0.9}{0.9967} & 0.5915 & -0.0498 & -0.0498 \\
  \hline
  Experiment ($\rho$) & \\ 
$\sigma_\alpha$ &  &  & -0.4630 & 0.4750 &\colorbox[rgb]{1,0.9,0.9}{~~0.9090} & \colorbox[rgb]{1,0.9,0.9}{~~0.9750} & \colorbox[rgb]{1,0.9,0.9}{~~0.9990} \\
($p_\alpha$) &  &  & (0.1667) & (0.2924) & \colorbox[rgb]{1,0.9,0.9}{(0.0014)} & \colorbox[rgb]{1,0.9,0.9}{(0.0000)} & \colorbox[rgb]{1,0.9,0.9}{(0.0000)} \\
 $\sigma_\beta$ &  &  &\colorbox[rgb]{1,0.9,0.9}{~~0.8510} &
  \colorbox[rgb]{1,0.9,0.9}{~~0.8400} & 0.3330 & 0.2250 & 0.0070 \\ 
 ($ p_\beta$) &  & & \colorbox[rgb]{1,0.9,0.9}{(0.0020)} & \colorbox[rgb]{1,0.9,0.9}{(0.0013)} & (0.1366) & (0.5280) & (0.8563) \\
  \hline 
 \end{tabular}
\end{table}

\paragraph{Parameter dependence predictions \label{sec:myopic}} 
Myopic response is the nature of human being. It play as the central principle in  behaviour game theory \cite{Behavioral2003}. By a simple myopic response strength estimation, we can estimate which mode should be qualitatively dominate (Details of the algorithm for estimation is shown in appendix section \ref{sec:myopic_algo}). 
Using the estimation, 
we can obtain the strength of the cycle, denoted as $L(a)$, 
for the 10 sub-spaces. 
The results are shown in the top panel in Table \ref{tab:ec_qualitative} (a full version is shown in Table \ref{tab:corr_nTE2} in appendix). 

Qualitatively, we can estimate 
projection of the strength of the response ($L(a)$) 
on the two mode $(\sigma_\alpha, \sigma_\beta)$ respectively
by calculate the correlation coefficients ($\rho$). 
The results are shown in the middle panel (labelled as Theory ($\rho$)) in Table \ref{tab:ec_qualitative}. The $\rho$ value have clear splitting between the various $a$. With this estimation, we believe that, the parameter selection is fit to answer above the research questions. 

Quantitatively, for various $a$ in game matrix, the eigenvector set is identical. The eigenvector set has only two orthogonal complex eigenvectors with their complex conjunction respectively. 
So, if a system can be decomposed by the eigen eigenvector $(\xi_a, \xi_b)$, 
by multi variable linear regression (MLR), we can learn how the weights depends on the parameter $a$ of the treatment.

\paragraph{Parameter independence predictions} As the components of the eigenvector is of high symmetry, the values of eigencycle is very limit, which can be $\pm \alpha$ and $\pm \beta$. So, between the 10 various 2-dimensional spaces, the observed eigencycle values can frequently be totally equal or totally oppose. This prediction is based on the eigen mode hypotheses: All parts of the eigenvector are not only oscillating with the same frequency, but also oscillating with constant phase difference. An example to interpret the invariant cross the 5 parameters are shown in appendix \ref{sec:example_invariant}

 The theoretical prediction of the invariant are shown in the upper triangular in Table \ref{tab:corr_e10xe10}, in which $(+)$ indicates the totally equal and $(-)$ indicates the totally oppose. 
Details of the algorithm for these theoretical results are shown in SI. In sum, for the 3rd RQ, we have follow prediction and its check method:

   \paragraph{Prediction 3} The motion observed in subspace have the relationship of  totally equal or totally oppose cross $a$; The relationship is predictable shown in Table \ref{tab:corr_e10xe10}, in which (+) indicates the equal and ($-$) indicates oppose. There are 20 points being expected to have the maximum
   correlation coefficients in significant. 

On verification,   
    this prediction can be verified 
    by the correlation coefficient ($\rho_{mn,m'n'}$ where the ($mn, m'n'$ are of the two 2-d subspace index) of the eigencycles values between the 10 2-dimensional subspace in experimental time series.

%

%

\subsection{Experimental observation} 
To identify the theoretical prediction shown in the logic chain, the test of of the long run mean and deviation of the strategy distribution, as well as the eigencycles. There are 5 experiment treatments which depends on the $a$ value. For each treatments, there has 6 sessions repeated. 
In each session having 20 minutes and 600 repeated rounds. The experiments were conducted during March 6 to April 3 at our laboratory.\footnote{We thanks San Lixia and Guo Hangyi for the help of the experiments carrying out.} Details of the experiment protocol see Method section. 
\subsubsection{Distribution and deviation} 
 Results of the mean and standard deviation do not reject the Nash equilibrium hypotheses. Notice that, the theoretical deviation depends on the population size. In our experiment, there are 6 human subjects participated, which can be deduced basing on the Nash equilibrium concept. Details of the results shown in Table \ref{tab:mean_x_trt_std_x_trt} in appendix.

\subsubsection{Eigencycle in experiment} 
Referring to the experimental eigencycle measurement (see Method), we can obtain the angular momentum results from the experiment time series for the five treatments respectively. The results are shown in Table \ref{tab:ec_t2e5} which labelled as $L^{\text{Tr}\cdot}$. 

In addtion, to identify the strength of the motion, we report the normal of the eigencycle amplitude, shown in the last row in Table \ref{tab:ec_t2e5}. 

\begin{table}[!ht] 
 \centering
 \caption{Theoretical eigencycle $\sigma$ and experimental observation $L$\label{tab:ec_t2e5}}
 \begin{tabular}{rrrrrrrr}
 \hline 
mn & $\sigma_a$ & $\sigma_\beta$ & $L^{\text{Tr}1}$ & $L^{\text{Tr}2}$ & $L^{\text{Tr}3}$ & $L^{\text{Tr}4}$ & $L^{\text{Tr}5}$ \\
\hline  
12 &    -0.3804 & 0.2351 & 0.3478 & -0.1060 & -0.2229 & -0.3899 & -0.4050 \\
13 &    -0.2351 & -0.3804 & -0.1601 & -0.4085 & -0.3906 & -0.2935 & -0.2057 \\
14 &    0.2351 & 0.3804 & 0.2400 & 0.5420 & 0.2760 & 0.2786 & 0.2349 \\
15 &    0.3804 & -0.2351 & -0.4278 & -0.0275 & 0.3375 & 0.4047 & 0.3758 \\
23 &    -0.3804 & 0.2351 & 0.4096 & -0.0319 & -0.1187 & -0.3265 & -0.4006 \\
24 &    -0.2351 & -0.3804 & -0.2292 & -0.3648 & -0.4537 & -0.2935 & -0.2081 \\
25 &    0.2351 & 0.3804 & 0.1675 & 0.2907 & 0.3495 & 0.2302 & 0.2038 \\
34 &    -0.3804 & 0.2351 & 0.4068 & 0.0591 & -0.1720 & -0.2655 & -0.4086 \\
35 &    -0.2351 & -0.3804 & -0.1572 & -0.4995 & -0.3373 & -0.3545 & -0.1977 \\
45 &    -0.3804 & 0.2351 & 0.4175 & 0.2363 & -0.3497 & -0.2803 & -0.3818 \\
 \hline 
 normal ampl.  & & & 8.6771 & 4.0952 & 2.6071 & 5.6943 & 11.5525 \\
 \hline 
\end{tabular}
\end{table}

\subsection{Verification by experiment} 
\subsubsection{Eigen mode selection} 
\paragraph{Weight of the eigen modes}
   Having the observed angular momentum vector $L^{Tr\cdot}$ from the experiment shown in Table \ref{tab:ec_t2e5}, we can calculate the weight of the $L^{Tr\cdot}$ over the eigencycle set vectors ($\sigma_\alpha, \sigma_\beta$). The calculation is carried by multiple linear regression analysis, which is used to assess the association between two independent variables ($\sigma_\alpha, \sigma_\beta$) and a single dependent variable, the outcome $L^{Tr\cdot}$. The multiple linear regression equation is as follows:
\begin{equation}
   L^{Tr\cdot} = c_0 + k_\alpha \sigma_\alpha + k_\beta \sigma_\beta.
\end{equation}
Herein, $k_\alpha$ and $k_\beta$ 
are the adjusted estimated regression coefficient 
that quantifies the association 
between the factors ($\sigma_\alpha$ and $\sigma_\beta$) 
and the outcome. 
The decomposition model have significant interpretation power 
when $p$ value close to 0 and $c_0$ 
does not deviate from 0 in significant. 

The results for the five treatments are shown in Table \ref{tab:c0k1k2}. 
We have also carried out the same calculate at experiment session level, in which there are 10 sample for each treatment, and the result shown in Table \ref{tab:k1andk2session}. Our interpretation of the results are following:
\begin{enumerate}
\item As response to RQ1, the null hypothesis is that the eigen mode does not exist or is indistinguishable. This null hypothesis can be rejected in significant by following results. 
\begin{itemize}
\item Results from the correlation coefficient ($\rho$) as well as the linear regression ($p$) value  between the eigencycle vectors of the theory and the experiment by treatment shown the bottom panel in Table \ref{tab:ec_qualitative}. In the Table, the \colorbox[rgb]{1,0.9,0.9}{highlight values} enhance the consistence between theory and experiment for visibility (see also the details in 1st-2nd column in Table \ref{tab:corr_nTE2}). 
\item Results from the $p$ value of the MLR by treatment show in Table \ref{tab:c0k1k2}
\item Results from the the $p$ value of the MLR by sessions (6th column in Table \ref{tab:k1andk2session}). 
\item Results from the statistical ttest results of the $p_{k_1}$ and the $p_{k_2}$ values by treatments, in which the samples comes from the $k_1$ and the $k_2$ of the MLR by sessions (the last two columns in Table \ref{tab:k1andk2session}). 
\end{itemize}
 These results illustrate that, the prediction 1 is supported. The dynamics observation $L^{Tr\cdot}$ from the all five treatment can be decomposed by the eigen modes in significant. 
 
\item For RQ2, the null hypothesis is that the mode selection is unpredictable. This can be rejected in significant. 
   
\begin{itemize}
  \item Evidence comes from experiment Tr2 and Tr4 reject the null hypotheses in significant. Because, the projection of $L^{Tr2}$ appears only in $\sigma_\beta$ ($p$=0.005, \textbf{ttest} on the deviation from 0, $N$=10 session); And $L^{Tr4}$ appears only in $\sigma_\beta$ ($p$=0.000, \textbf{ttest} on the deviation from 0, $N$=10 session). The statics can be obtained  in Table \ref{tab:k1andk2session} in the appendix.
  \item Supporting evidence also comes from experiment Tr1 and Tr3, of which theoretical expectation is that, the weight $|k_\alpha|$ and $|k_\beta|$ are same. 
  Both Tr1 ($p$ = 0.090, \textbf{ttest} on the null hypothesis $|k_\alpha| = |k_\beta|$, $N$=10 session) and Tr3 ($p$= 0.386, \textbf{ttest} on the null hypothesis $|k_\alpha| = |k_\beta|$, $N$=10 session) do not rejected the theoretical in significant. The statics can be obtained base on the data shown in Table \ref{tab:k1__k2sess} in the appendix. 
\end{itemize} 
\end{enumerate}

\begin{table}
  \caption{Multi eigen mode weight}
  \label{tab:c0k1k2}
  \centering
  \begin{tabular}{cr|rrrrr}
\hline
Treatment  & $a$~~~~  & $c_0$ & $k_\alpha$ & $k_\beta$ &$ \rho$~~~ & $p$~~~~\\ 
\hline
Tr1  &  $-$4.2360 & $-$0.0000 & $-$0.5967 & 0.9229 & 0.9884 & 0.0000 \\
Tr2   &  $-$0.6180 & 0.0000 & 0.2214 & 0.4403 & 0.8924 & 0.0004 \\
Tr3  &  0.2360 & 0.0000 & 0.2927 & 0.1345 & 0.9337 & 0.0001 \\
Tr4  &  1.6180 & $-$0.0000 & 0.7070 & 0.1159 & 0.9716 & 0.0000 \\
Tr5  &  4.2360 & $-$0.0000 & 1.4666 & $-$0.0943 & 0.9979 & 0.0000 \\
\hline
 \end{tabular}
\end{table}

\subsubsection{Invariant in eigen modes}
In theory, there exist invariant between the cycles in the subspace. 
The motion observed in some 2d sub-spaces 
can be totally equal or totally oppose independent on $a$; 
The prediction is predictable shown in the up triangle in the top panel in Table \ref{tab:corr_e10xe10}. 
In the up triangle, the $(+)$ and $(-)$ indicate the equal and oppose, respectively; If no correlated, the cell remain blank.  
With the total 50 sessions data, this prediction can be verified by 
the correlation coefficient $\rho$ of 
the eigencycles values between the 10 
2-dimensional subspace in experimental time series.
 
 As a clean up presentation, Table \ref{tab:corr_e10xe10} reports the correlation coefficient $\rho$ of observed eigencycle between the 10 in pairs (Full results is shown in Table \ref{tab:corr_e10xe10_full} in appendix). 
 Notice that,  in $\rho_{35,14}$, the low script ($35, 14$) are of the two 2-d subspace index, ($x_3,x_5$) and ($x_1,x_4$), respectively; and $x_.$ is the proposition of the $x_.$ strategy in games. 
 There are totally 45 independent test points. The bottom panel reports the $p$ value of linear regression to verified  the dependence of the paired 50 observations. So call clean up means 
 that we select the 20 maximum $|\rho|$ value to fill the top panel, meanwhile the 20 minimum $p$ value to fill the bottom panel in Table \ref{tab:corr_e10xe10}  from the Table \ref{tab:corr_e10xe10_full}.      

As result, referring to the reported $\rho$ and $p$ values, none of the 45 observed check points violates the theoretical prediction. This is the answer to the RQ3, there exists invariant between the subspace insignificant ($p$=0.0000, \textbf{Wilcoxon Sign Rank Test}, $N$=45 check points).

\begin{table} 
 \centering
 \caption{correlation coefficient of observed eigencycle between 10 2-dimensional subspace \label{tab:corr_e10xe10}}
 \footnotesize
\begin{tabular}{r|rrrrrrrrrr} 
\hline 
 & $L^{12}$ & $L^{13}$ & $L^{14}$ & $L^{15}$ & $L^{23}$ & $L^{24}$ & $L^{25}$ & $L^{34}$ & $L^{35}$ & $L^{45}$ \\
 \hline 
$\rho$ \\
 \hline  
 $L^{12}$ &  1& & & 
 ($-$)&  (+)& & & (+)& & (+)\\
$L^{13}$ &  & 1& ($-$)& & & (+)& ($-$)& & (+)&\\
$L^{14}$ &  & -0.5832& 1& & & ($-$)& (+)& & ($-$)&\\
$L^{15}$ &  -0.9101& & & 1& ($-$)& & & ($-$)& & ($-$)\\
$L^{23}$ &  0.9380& & & -0.8719& 1& & & (+)& & (+)\\
$L^{24}$ & & 0.5599& -0.6210& & & 1& ($-$)& & (+)&\\
$L^{25}$ &  & -0.5855& 0.5656& & & -0.7145& 1& & ($-$)&\\
$L^{34}$ &  0.8823& & & -0.8455& 0.9427& & & 1& & (+)\\
$L^{35}$ &  & 0.7611& -0.4853& & & 0.5719& -0.6138& & 1& \\
$L^{45}$ &  0.8826& & & -0.9260& 0.8826& & & 0.9102& & 1\\
 \hline 
$p$  \\
 \hline 
$L^{12}$ &     0 &  &  &  &  &  &  &  &  &  \\
$L^{13}$ &      & 0 &  &  &  &  &  &  &  &  \\
$L^{14}$ &     & 0.0000 & 0 &  &  &  & &  &  &  \\
$L^{15}$ &  0.0000 &  &  & 0 &  &  &  &  &  &  \\
$L^{23}$ & 0.0000 &  &  & 0.0000 & 0 &  &  &  &  &  \\
$L^{24}$ &   & 0.0000 & 0.0000 &  &  & 0 &  &  &  &  \\
$L^{25}$ &   & 0.0000 & 0.0000 &  &  & 0.0000 & 0 &  &  &  \\
$L^{34}$ & 0.0000 &  &  & 0.0000 & 0.0000 &  &  & 0 &  &  \\
$L^{35}$ &   & 0.0000 & 0.0004 &  &  & 0.0000 & 0.0000 &  & 0 &  \\
$L^{45}$ & 0.0000 &  &  & 0.0000 & 0.0000 &  &  & 0.0000 &  & 0 \\
\hline
  \end{tabular}
\end{table}

\section{Discussion}
We have illustrated the eigen mode existence as well as the eigen mode selection in laboratory human subject game experiment. The eigen mode selection, which identified by eigencycle and angular momentum measurement,  can be qualitatively  
predicted by two theoretical approaches: the replicator dynamics and the principle of the myopic best responses of human natural on decision making. 

To obtain the eigen mode selection is not an accident, when we refer to the cycles in elementary games published near 2014 \cite{dan2014, wang2014social, wang2014}, as well as  recent finding and confirmation in laboratory game experiments: 
\begin{enumerate}
  \item By the end of 2020, in the O'Neill game experiment reported in 1987 \cite{ONeill1987}, the theoretical prediction of
  the eigensystem has been confirmed by eigencycle measurement. 
  In the O'Neill game experiment reported in 2001 \cite{Binmore2001Minimax} and in 2013 \cite{Yoshitaka2013Minimax},
  the dynamics structures in various setting are consistence in high accuracy in 2020 \cite{WY2020,2021Qinmei}
  \item In a five strategy one population game experiment, 
  using the same approach, the regularity of high dimensional cycle
  is confirmed, in 2021. \cite{2021Qinmei} 
  \item In a four strategy one population game experiment, 
  using the same approach again, the regularity of high dimensional cycle
  is confirmed by two controlled parameter, too, in 2021. \cite{2021Shujie}
  \item It has been notice that, 
  borrowing the pole assignment approach in state dependent closed-loop feedback control theory in modern control theory, 
  the game dynamics structure control by design is realisable in laboratory game experiments in 2022. \cite{wang2022} 
\end{enumerate}
Among these, only the O'Neill game (designed by O'Neill 1987 \cite{ONeill1987}) has 3 pairs of complex eigenvalue/eigenvector. It has been notice that, among the three pairs, the low eigenvalue has significant influence in the dynamics structure, which has already reach the eigen mode selection issue (see Eq. 6.5 in the master thesis \cite{2021Qinmei}). 
So, it is not a surprise for us to pay attention on multi mode selection. 

At its root, this study is follow the workflow (logic chain) 
which has been existed since 1978 when the game dynamics established \cite{taylor1978evolutionary}. At that time, Jacobian matrix (character matrix) had been there before 1850. In 2020 \cite{WY2020}, the finding of the dynamics structure (cycling spectrum) in O'Neill game reported in 1987 is only an application of the long existed workflow. 
Might why finding such simple and clear pattern need 
so long waiting is a puzzle in the history of studying game dynamics theory.    

This study add knowledge to the existed workflow led by \cite{WY2020},  by provides the myopic response estimation to predict the eigen mode selection (see Table \ref{tab:ec_qualitative} and its related explanation as example). Although the estimation is qualitative approach, but it highlight that, in game dynamics theory, the eigen mode selection issue is not totally in dark.   

On further, the consistency of the theoretical approaches has to be investigated.\footnote{Wang Zhijian thanks Zhao Yuqi from Chu Cochen Honors College of Zhejiang University, as well as Guo Hangyi from School of Computer Science and Engineering of Zhejiang University for helpful discussion at this point.}  
There are two approaches
\begin{itemize}
\item The eigenvalues from replicator dynamics 
in Eq. (\ref{eq:eigenvalue}), considering the impact of eigenvalue on the angular momentum measurement, which was proved in mathematics in \cite{WY2020} (see Appendix 2: Invariant between eigencycle and the angular momentum in \cite{WY2020}).  
\item   The myopic response strength estimation approach in section \ref{sec:myopic} and shown in Table \ref{tab:ec_qualitative}, which play as the central principle in behaviour game theory \cite{Behavioral2003}. 
\end{itemize}
 in this report to predict 
the eigen mode selection.  Their predictions  are supported 
in statistics significant by the experiments data in this study. However, none of the two approaches straightly predicts the weight of the mode selected. This is an obvious challenge to game theorist.


\section{Appendeix}
\subsection{Technology Notes}
\subsubsection{The consequence of the oppose a}
When oppose $a$ to $-a$, one may think the game well remain same by trans the s1 strategy with the s5 strategy, the s2 strategy with 3 strategy, and then the matrix should be: 
\begin{equation}
     {A}^\dag =  \left(\begin{array}{rrrrr}
        0 & -a & -1 & 1 & a\\
        a & 0 & -a & -1 & 1\\
        1 & a & 0 & -a & -1\\
        -1 & 1 & a & 0 & -a\\
        -a & -1 & 1 & a & 0
        \end{array}\right)
\end{equation} 
This matrix appears similar to the original, but actually is not equivalent because of -1 and 1 elements in the matrix. So, the theoretical results of the Treatment 1 ($a=-4.236$) can be different from the Treatment 5 ($a=+4.236$).

\subsubsection{Details of the eigenvector structure}
\begin{figure}[ht!]
\centering 
\includegraphics[scale=0.6]{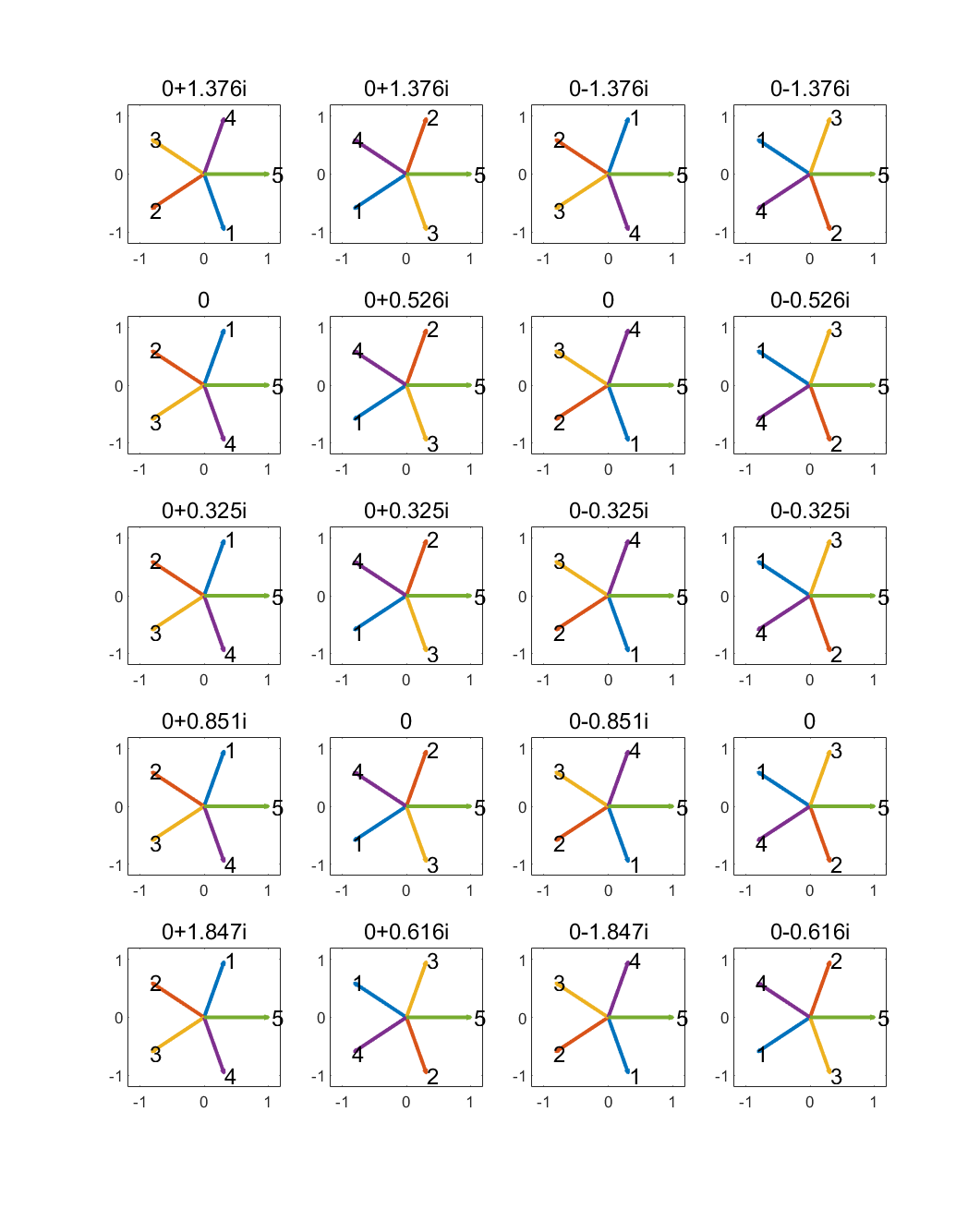}
    \caption{Geometric presentation of the eigenvector set. From top to bottom are by the order of $a$, from -4.236 to 4.236. The titles of each sub figure is the associate eigenvalue of the eigenvector.  \label{fig:evec5x4}}
\end{figure}

The eigenvector structure of the game play cruel role in this study. 
Besides the unique equilibrium (0.2, 0.2, 0.2, 0.2, 0.2) for various $a$, the eigenvector structure are of high symmetry, which has only two independent patterns, for various $a$ too. Details of the pattern for
various $a$ are shown in Fig. \ref{fig:evec5x4}. 

It is obviously that, among these pattern, only the two $v^\alpha$ and $v^\beta$ shown in Fig. \ref{fig:eig_vec} in main text are independent.
the remains are of the complex conjunction of these two vectors. 

\subsubsection{The 5 parameter selection \label{app:a_select}}
 It is easy to notice that, the manifold satisfy central manifold theorem requirement: 
$$\Re (\lambda_i) = 0 ~~~~ i \in \{1,2,3,4,5\}$$
At the same time, when $a=-0.618$ $$\lambda_2 = \lambda_4 = 0$$
and when $a=1.618$ $$\lambda_3 = \lambda_5 = 0$$
Meanwhile, when $$a=\left(\begin{array}{c}  - \sqrt{5} - 2\\ \sqrt{5} - 2 \end{array}\right) = \left(\begin{array}{r} -4.236\\  0.236 \end{array}\right)$$
the system has unique imaginary part, 
$$ \begin{array}{c} |\Im(\lambda_i)| \end{array} =\left(\begin{array}{c} \sqrt{ - 2\, \sqrt{5} - 1}   \\ \sqrt{ 2\, \sqrt{5} - 1} \end{array} \right)  ~~~~ i \in \{2,3,4,5\}. $$
Having same imaginary part indicates having same frequency of cycles. 

For the symmetry consideration, we choose the oppose the maximum value above is 4.236, which is chose as the last parameter.  In sum, the parameters selected are 
$$a= \big[ -4.236,	-0.618,	0.234,	1.618,	4.236 \big].$$

%
%
%
%
%
%
%
%
%
%


\subsubsection{Measurement of experimental angular momentum}

\paragraph{The measurement for the cycle in the subspace: }

According to the theoretical eigencycle set decomposition approach,
we can measure the cyclic angular momentum
~\footnote{This measurement is the signed area of
the triangle $\Delta_{[O, x(t), x(t+1)]}$ in the ($m,n$) two-dimensional subspace.
For each transition from $x(t)$ to $x(t+1)$, referring to $O$,
the angular momentum is twice the signed area of the triangle.
We suggest using the angular momentum
because it contains
the mass $m$ as a parameter,
which may be compatible with the population size $N$
as the variable in further investigations of game dynamics. }
in each of the two-dimensional subspace,
indicated by the eigencycle $\Omega^{(mn)}$, separately.
The angular momentum $L^{(mn)}_E$ \cite{wang2017} can be expressed by the following formula: \\
\begin{equation}\label{eq:exp_am} L^{mn}_E=\frac{1}{N-1}\sum_{t=1}^{N-1}\left(x(t)-O\right) \times \left(x(t\!+\!1)-x(t)\right)
\end{equation}

\begin{itemize}
\item $L^{(mn)}_E$ represents the average value of
the accumulated angular momentum over time;
the subscript $mn$ indexes the two-dimensional $(x_m,x_n)$ subspace;
\item $N$ is the length of the experimental time series, that is,
the total number of repetitions of the repeated game experiments;
\item $O$ is the projection of the Nash equilibrium at the subspace $\Omega^{(m,n)}$;
\item $x(t)$ is a two-dimensional vector at time $t$,
which can be expressed as $(x_m(t),x_n(t))$, $x(t+1)$ is a two-dimensional vector at time $t+1$, and
\item $\times$ represents the cross product between two two-dimensional vectors.
\end{itemize}
 
 Interpretation of the measurement

\subsubsection{Theoretical eigencycle set\label{app:def_ec}}

\paragraph{Definition} 
We call the circle constructed by the
two components $(\eta_m,\eta_n)$ within one normalized eigenvector
${\xi}_i = (\eta_1, ..., \eta_m, ..., \eta_n, ... \eta_s)$
the eigencycle, marked as $\sigma^{(mn)}$ and calculated as follows:
\begin{equation}\label{eq:the_ecyc}
 \sigma^{(mn)}=\pi \cdot ||\eta_m|| \cdot ||\eta_n||
               \cdot  \sin\big (\arg(\eta_m)-\arg(\eta_n)\big ),
\end{equation}
where $m$ and $n$ are the abscissa and the ordinate dimension
of the two dimension subspace, respectively.
$||\eta_m||$ and $\arg(\eta_m)$ indicate
the amplitude and the phase angle
of the $\eta_m$, respectively.
$\sigma^{(mn)}$ can determine the direction and amplitude of the eigencycle.
An alternative and equivalent presentation of the eigencycle is
\begin{equation}\label{eq:the_ecyc2}
       \sigma^{(mn)}=\pi \cdot \big (\Re(\eta_m)\Im(\eta_n)
                                    - \Re(\eta_n)\Im(\eta_m)\big ),
\end{equation}
wherein the $\Re(\eta_m)$ is the real part of
the complex number $\eta_m$ and
the $\Im(\eta_m)$ is the imaginary part.

\paragraph{Eigencycle set values of the game} 
According to this formula, for these game, the column $\sigma^\alpha$ and  $\sigma^\beta$ in 
Table \ref{tab:ec_t2e5} lists the eigencycle set values
of the two unique eigenvectors of the replicator dynamics.\\
\begin{flushleft}
\textbf{Interpretation}
\end{flushleft}
\begin{itemize} 
\item \textbf{Eigencycle set:} The eigencycle set, as a vector denoted by $\Omega^{(mn)}_{\xi_k}$, is defined to represent a set of $N(N-1)/2$ eigencycle elements. The subscript is the normalized eigenvector $\xi$ indexed by $k$, which generates this eigencycle set. The superscript $(mn)$ is the index of the two-dimensional subspace, where the elements (eigencycles) of the set are located. $(mn)$ is defined as follows: \{$\{m,n\} \in \{1,2,...,n\} \cap (m < n)$\}. In this study, the assignment order is $m$ from 1 to $N$, and then $n$ from 2 to $N$.
\item \textbf{Subspace set:} The subspace set, denoted as $\Omega^{(mn)}$, has $N(N-1)/2$ eigencycle elements. The superscript $(mn)$ is the index of the two-dimensional subspace. Again, $(mn)$ is defined as follows: \{$\{m,n\} \in \{1,2,...,n\} \cap (m < n)$\}.  
\item \textbf{Geometric presentation:}  An eigencycle only depends on the internal components $\eta_m$ and $\eta_n$ of the normalized eigenvector. As both of the components' amplitude and phase difference is fixed and not arbitrary, so the geometric pattern set is fixed. In geometry, an elements of the eigencycle set is similar to a Lissajous cycle as shown in figure \ref{fig:eig_lissa}. 
\end{itemize}


\subsubsection{Equivalent of eigencycle and angular momentum} \label{app:sigma_L}
In mathematics, we can strictly prove that, the rate between the angular momentum at any time moment in any two subspace equals to  the rate between eigencycle values of the two subspace. In a time series from a real system, the angular momentum of two observation is measurable, so the eigencycle concept can be verified in real system.

The invariant (Equivalent) between
the eigencycle and the angular momentum is important in this study.
Because, only having this invariant, we can using eigencycle as predictor
for experimental angualr momentum.
Here, the eigencycle as predictor can be obtained
by the eigenvector from dynamics equation, and
the angular momentum can he obtained in time series.
 
Main idea to prove the invariant is
to calculate the angular momentum of the eigencycle explicitly. Details of the deduction are shown in \cite{WY2020}.
The main results applied in this study are following two propositions.  
 
 \paragraph{Proposition 1}  
 For a given eigenvector $\xi$, in all of the two dimension sub-spaces $\Omega^{(m,n)}$
 at any given $t$, the rate between the instantaneous angular momentum value and eigencycle value is equal,  \\
\begin{equation}
    \frac{L_{mn}(t)}{\sigma^{(mn)}} = \textbf{C}(t) ~~~~ \forall (m,n) \in (1,2,...,N)
\end{equation}
Such that,  for various subspace $(m,n)$ and $(m',n')$,
the average accumulated angular momentum
$L_{mn}$ divided by its $\sigma^{(mn)}$ is  \textbf{invariant},
\begin{equation}
   \frac{1}{t_1 - t_0} \sum_{t=t_0}^{t_1}\frac{L_{mn}(t)}{\sigma^{(mn)}}
   = \frac{1}{t_1 - t_0} \sum_{t=t_0}^{t_1}\frac{L_{m'n'}(t)}{\sigma^{(m'n')}}
   = \textbf{C}  ~~~~ \forall (m,n;m',n') \in (1,2,...,N)
\end{equation}
   So, by take $\sigma^{(mn)}$ as predictor (independent variable) and take $L_{mn}$ as the observation (dependent variables), the a linear regression line, with truncation term being 0,
   is expected.

 \paragraph{Proposition 2}  
 Interfere of two components from two different eigenvector is vanish,
 when following condition hold:\\
    \begin{enumerate}
      \item White noise condition: after a random interrupt, the initial phase
            of the two eigenvector distribute between [$-\pi, \pi$] equally;
      \item Sufficient noise shock: there are sufficient times of
            the interrupt;
      \item Long time limit: there are sufficient long time  of the time series.
    \end{enumerate}

\subsubsection{Mode of angular momentum} 
We define mode of angular momentum as, 
$$\Big|L\Big| =  \Big(\sum_{i=1}^{10} L_{mn}^2\Big)^{1/2}~~~~~~~~~ (m,n) \in \{1,2,3,4,5\} \cap m<n$$

\subsubsection{Algorithm for the myopic response strength estimation \label{sec:myopic_algo}} 

 Following estimation is not an approach for general game. This is what we hope to emphasis.   
 
This estimation takes the advantage of the symmetry of the game. So called symmetry, are of the equilibrium and the eigenvector structure, as well as the payoff matrix and Jaobian matrix. 

At the same time, we assume that, In long run,  statistically, the deviation of the each strategy are symmetrical (equal). 

Now, assume that, in a population with huge number of agents system $S$. The system is exactly at Nash equilibrium, and in which, we add two agents A and B, and both A and B play $x_1$ strategy.  This is the initial condition. 

Now, only A has opportunity to make a response. Its response will lead to the change of probability of $x_2, x_3, x_4, x_5$, which have consequence on angular momentum measurement.  

When A makes a response,  the observed velocity of $\dot{x}_2$ in the sub space $mn=(1,2)$ is close to
$$\dot{x}_2 \simeq J_{21}\cdot \delta(x_1).$$
then the expected angular momentum measured in in the sub space $(mn)=(1,2)$ is 
$$L_{1}  \simeq \delta(x_1) \times \dot{x}_2 \simeq J_{21}\cdot \delta(x_1)^2.$$
Herein, $J$ is the Jacobian (character matrix) of the dynamics as shown in Eq. \ref{eq:J} in main text. Similarly, if the added two agents are playing $x_2$, we can expect that
$$L_{2}  \simeq - \delta(x_2) \times \dot{x}_1 \simeq - J_{12}\cdot \delta(x_2)^2.$$
Here, the symbol $(-)$ is added because of the definition of angular momentum. 

As the deviation of $\delta(x_1)$ and  $\delta(x_2)$ are equal in long run, statistically, these two factors have same contribution on $L$.

At the same time, in this study case $J_{ij} = -J_{ji}$ for all $(i,j)=(1,2,3,4,5) \cap (i>j)$, so the 
contribution of the two terms are equal, 
and propitiation to $J_{12}$. 

In this study case, 
there two constant ($\pm 1$) in $J$ 
which are applied as invariant. 
As estimation, we set step function $f$ as
\begin{itemize}
\item ~~~~2, ~~if $J_{ij} > 1$
\item ~~$-$2, ~~if $J_{ij} < -1$
\item ~~1/2, ~~if $0 < J_{ij} <  1$
\item $-$1/2, ~~if $-1 < J_{ij} < 0$
\end{itemize}

Similarly we can assume the deviation appear at $x_2, x_3, x_4, x_5$ respectively for various game parameter $a$.  Then the expected angular momentum set  for each $a$ in Table \ref{tab:ec_qualitative} can be obtained.  

Then, by evaluating the projection (correlation coefficient $\rho$) of these angular momentum set (scale vector) over the eigencycle set (scale vector), we provide the theoretical predictions of the eigen mode decomposition weight (eigen mode selection) for various game with the parameter $a$.

\subsubsection{An example of the invariant cross the 5 parameters \label{sec:example_invariant}}
We show an example, using the observations from the subspace (12) and (15) , to explain the prediction of the parameter independence.  Suppose a motion $x(t)$ can be decomposed as
$$x(t) = C_a \xi_a + C_b \xi_b $$
in which, $C_i$ is the weight (or contribution) of the eigenvector $i$. For different treatment the weight could differ. So, the theoretical composed eigencycle $\Omega$ in the 2-d subspace (12) can be expressed as  
$$\Omega^{12} = C_a^2 \sigma_a^{12} + C_b^2 \sigma_b^{12}; $$ 
At the same time, the composed eigencycle $\Omega$ in the 2-d subspace (15) is 
$$\Omega^{15} = C_a^2 \sigma_a^{15} + C_b^2 \sigma_b^{15}; $$ 
as $\sigma_1^{12}=-\sigma_1^{15}$ and  
$\sigma_2^{12}=-\sigma_2^{15}$. 
So, disregarding the combination of ($C_1, C_2$) for various treatment in theory, which is instinct dynamics property of the game. we can always have 
$$\Omega^{12} = - \Omega^{15}. $$ 
Then observations from the subspace (12) and (15) are of the totally oppose relation, which is of parameter independence.

In the same way, we deduce the theoretical expectation of the parameter independence relationship for each pair of the 10 subspace.

\begin{table} 
 \centering
 \caption{Experiment  $L$ in 10 sub space By Treatment (\textbf{Tr1-Tr5}) and Sessions (\textbf{1-10}) \label{tab:ezp_sess}}
 \small
 \begin{tabular}{r|r|rrrrrrrrrr|}
 \hline 
 &mn&	\textbf{1}&	\textbf{2}&	\textbf{3}&	\textbf{4}&	\textbf{5}&	\textbf{6}&	\textbf{7}&	\textbf{8}&	\textbf{9}&	\textbf{10}\\
 \hline  
\textbf{Tr1} &\textbf{12}&	3.733&	1.744&	1.6&	1.806&	7.694&	0.556&	4.783&	1.994&	4.194&	2.078\\
& \textbf{13}&	-1.017&	-3.944&	-2.628&	-0.844&	-0.456&	-0.233&	-2.439&	-1.072&	-1.067&	-0.189\\
& \textbf{14}&	-0.539&	6.089&	3.95&	0.967&	0.072&	2.394&	2.783&	1.311&	1.567&	2.228\\
& \textbf{15}&	-2.178&	-3.889&	-2.922&	-1.928&	-7.311&	-2.717&	-5.128&	-2.233&	-4.694&	-4.117\\
& \textbf{23}&	2.889&	1.922&	4.078&	3.739&	7.906&	1.6&	3.5&	3.322&	4.683&	1.9\\
& \textbf{24}&	-0.3&	-5.628&	-3.594&	-3.061&	-1.206&	-1.828&	-1.639&	0.233&	-1.906&	-0.961\\
& \textbf{25}&	1.144&	5.45&	1.117&	1.128&	0.994&	0.783&	2.922&	-1.561&	1.417&	1.139\\
& \textbf{34}&	3.089&	2.144&	3.911&	5.922&	7.167&	1.656&	1.778&	3.55&	4.106&	1.972\\
& \textbf{35}&	-1.217&	-4.167&	-2.461&	-3.028&	0.283&	-0.289&	-0.717&	-1.3&	-0.489&	-0.261\\
& \textbf{45}&	2.25&	2.606&	4.267&	3.828&	6.033&	2.222&	2.922&	5.094&	3.767&	3.239\\
 \hline  
\textbf{Tr2}& \textbf{12}&	-1.267&	-1.45&	0.161&	-0.572&	2.567&	-3.339&	1.767&	-2.139&	0.133&	-0.2\\
& \textbf{13}&	-1.522&	-3.106&	-3.239&	0.35&	-4.217&	-0.783&	1.333&	-2.833&	-1.639&	-1.072\\
& \textbf{14}&	1.417&	4.039&	3.644&	0.878&	3.367&	2.383&	1.45&	2.283&	2.017&	0.717\\
& \textbf{15}&	1.372&	0.517&	-0.567&	-0.656&	-1.717&	1.739&	-4.55&	2.689&	-0.511&	0.556\\
& \textbf{23}&	-1.506&	-1.6&	0.906&	-1.856&	3.189&	-2.667&	1.928&	0.639&	0.061&	-0.4\\
& \textbf{24}&	-0.372&	-2.983&	-2.739&	0.2&	-2.561&	-1.972&	0.683&	-3.883&	-0.506&	-0.806\\
& \textbf{25}&	0.611&	3.133&	1.994&	1.083&	1.939&	1.3&	-0.844&	1.106&	0.578&	1.006\\
& \textbf{34}&	-1.811&	-0.8&	0.744&	-0.056&	3.194&	-1.5&	2.117&	0.561&	0.406&	-0.433\\
& \textbf{35}&	-1.217&	-3.906&	-3.078&	-1.45&	-4.222&	-1.95&	1.144&	-2.756&	-1.983&	-1.039\\
& \textbf{45}&	-0.767&	0.256&	1.65&	1.022&	4&	-1.089&	4.25&	-1.039&	1.917&	-0.522\\
 \hline  
\textbf{Tr3}& \textbf{12}&	0.422&	-1.4&	0.661&	0.933&	-1.283&	-2.528&	-1.139&	-0.122&	-0.956&	-0.4\\
& \textbf{13}&	-0.611&	-0.8&	-2.422&	-1.556&	-1.894&	-0.544&	1.144&	-1.778&	1.128&	-2.85\\
& \textbf{14}&	-0.156&	0.778&	1.628&	0.744&	2.861&	1.344&	-0.656&	-0.283&	-0.717&	1.65\\
& \textbf{15}&	0.344&	1.422&	0.133&	-0.122&	0.317&	1.728&	0.65&	2.183&	0.544&	1.6\\
& \textbf{23}&	0.578&	-0.956&	-0.417&	0.872&	0.306&	-3.1&	-0.467&	0.122&	-0.361&	0.328\\
& \textbf{24}&	0.089&	-3.017&	-0.756&	-1.439&	-2.828&	-0.517&	-0.822&	-0.8&	0.017&	-1.756\\
& \textbf{25}&	-0.244&	2.572&	1.833&	1.5&	1.239&	1.089&	0.15&	0.556&	-0.611&	1.028\\
& \textbf{34}&	-0.35&	-0.783&	-1.394&	0.717&	-0.828&	-2&	1.061&	-0.144&	-0.067&	-0.694\\
& \textbf{35}&	0.317&	-0.972&	-1.444&	-1.4&	-0.761&	-1.644&	-0.383&	-1.511&	0.833&	-1.828\\
& \textbf{45}&	-0.417&	-3.022&	-0.522&	0.022&	-0.794&	-1.172&	-0.417&	-1.228&	-0.767&	-0.8\\
 \hline  
\textbf{Tr4}& \textbf{12}&	-3.817&	-2.472&	-1.628&	-1.161&	-3.106&	-2.922&	0.494&	-4.067&	-1.172&	-2.35\\
& \textbf{13}&	-2.322&	-3.5&	-1.811&	-1.333&	-1.489&	-1.039&	0.556&	-2.656&	-2&	-1.117\\
& \textbf{14}&	0.883&	2.739&	2&	0.306&	4.389&	0.733&	0.256&	3.139&	-0.733&	2.156\\
& \textbf{15}&	5.256&	3.233&	1.439&	2.189&	0.206&	3.228&	-1.306&	3.583&	3.906&	1.311\\
& \textbf{23}&	-2.922&	-2.444&	-2.767&	-1.089&	-1.744&	-2.672&	1.2&	-3.478&	-1.272&	-1.406\\
& \textbf{24}&	-1.328&	-2.011&	-1.65&	-0.839&	-4.089&	-0.678&	-1.183&	-2.517&	-0.783&	-1.633\\
& \textbf{25}&	0.433&	1.983&	2.789&	0.767&	2.728&	0.428&	0.478&	1.928&	0.883&	0.689\\
& \textbf{34}&	-1.767&	-1.956&	-1.272&	-0.778&	-1.733&	-2.033&	2.394&	-2.9&	-2.956&	-2.117\\
& \textbf{35}&	-3.478&	-3.989&	-3.306&	-1.644&	-1.5&	-1.678&	-0.639&	-3.233&	-0.317&	-0.406\\
& \textbf{45}&	-2.211&	-1.228&	-0.922&	-1.311&	-1.433&	-1.978&	1.467&	-2.278&	-4.472&	-1.594\\
 \hline  
\textbf{Tr5}& \textbf{12}&	-5.533&	-4.8&	-1.511&	-5.172&	-2.8&	-6.739&	-0.622&	-4.972&	-6.306&	-8.333\\
& \textbf{13}&	-4.344&	-5.461&	-1.517&	-3.333&	-0.211&	-1.2&	-3.006&	-2.722&	-0.467&	-1.506\\
& \textbf{14}&	2.178&	5.017&	0.661&	1.45&	4.306&	3.006&	3.639&	2.589&	1.144&	3.15\\
& \textbf{15}&	7.7&	5.244&	2.367&	7.056&	-1.294&	4.933&	-0.011&	5.106&	5.628&	6.689\\
& \textbf{23}&	-5.394&	-4.383&	-4.367&	-4.606&	-4.022&	-3.822&	-1.661&	-4.083&	-7.3&	-6.644\\
& \textbf{24}&	-4.928&	-3.517&	-0.772&	-2.156&	-0.894&	-3.839&	0.733&	-3.328&	-0.383&	-4.961\\
& \textbf{25}&	4.789&	3.1&	3.628&	1.589&	2.117&	0.922&	0.306&	2.439&	1.378&	3.272\\
& \textbf{34}&	-4.339&	-4.05&	-3.1&	-7.717&	-4.956&	-3.739&	-3.133&	-5.217&	-5.583&	-5.372\\
& \textbf{35}&	-5.4&	-5.794&	-2.783&	-0.222&	0.722&	-1.283&	-1.533&	-1.589&	-2.183&	-2.778\\
& \textbf{45}&	-7.089&	-2.55&	-3.211&	-8.422&	-1.544&	-4.572&	1.239&	-5.956&	-4.822&	-7.183\\ 
 \hline  
 \end{tabular}
\end{table}


\begin{table}  
 \centering
 \caption{ Mean and STD \label{tab:mean_x_trt_std_x_trt}}
 \begin{tabular} 
     	{c|ccccc|ccccc}
 \hline  
   &   	 $p(x_1)$ & $p(x_2)$ & $p(x_3)$ & $p(x_4)$ & $p(x_5)$ & $Std(x_1)$ & $Std(x_2)$ & $Std(x_3)$ & $Std(x_4)$ & $Std(x_5)$ \\
 \hline  
  \textbf{Tr1} &   	 0.194 & 0.209 & 0.188 & 0.198 & 0.210 & 0.156 & 0.162 & 0.153 & 0.162 & 0.163 \\
  \textbf{Tr2} &       	 0.213 & 0.215 & 0.185 & 0.188 & 0.199 & 0.158 & 0.162 & 0.157 & 0.158 & 0.158 \\
  \textbf{Tr3} &       	 0.216 & 0.216 & 0.181 & 0.188 & 0.198 & 0.162 & 0.168 & 0.152 & 0.153 & 0.157 \\
  \textbf{Tr4} &       	 0.212 & 0.201 & 0.183 & 0.196 & 0.207 & 0.161 & 0.162 & 0.156 & 0.157 & 0.161 \\
  \textbf{Tr5} &       	 0.209 & 0.196 & 0.195 & 0.194 & 0.207 & 0.163 & 0.158 & 0.159 & 0.157 & 0.157 \\
 \hline  
 \end{tabular}
\end{table}


\begin{table}  
 \centering
 \caption{Mean and STD by sessions in experiment \label{tab:[mean_x_ses_std_x_ses]}}
 \begin{tabular} 
     	{cc|ccccc|ccccc|}  
  \hline  
      ($a$)&	session&
     	$p(x_1)$ & $p(x_2)$ & $p(x_3)$ & $p(x_4)$ & $p(x_5)$ & $ s(x_1)$ & $s(x_2)$ & $ s(x_3)$ & $ s(x_4)$ & $ s(x_5)$ \\
 \hline  
     	  \textbf{Tr1}  & 1 & 0.185 & 0.209 & 0.214 & 0.174 & 0.218 & 0.142 & 0.158 & 0.139 & 0.139 & 0.152 \\
     	 (-4.236) & 2 &0.221 & 0.227 & 0.159 & 0.199 & 0.194 & 0.170 & 0.176 & 0.149 & 0.178 & 0.163 \\
     	 & 3 & 0.185 & 0.221 & 0.187 & 0.206 & 0.201 & 0.158 & 0.174 & 0.152 & 0.171 & 0.167 \\
     	  & 4 &0.190 & 0.175 & 0.224 & 0.209 & 0.202 & 0.153 & 0.140 & 0.167 & 0.178 & 0.165 \\
     	  & 5 &0.190 & 0.210 & 0.207 & 0.191 & 0.202 & 0.165 & 0.175 & 0.175 & 0.159 & 0.173 \\
     	 & 6 & 0.196 & 0.219 & 0.160 & 0.241 & 0.184 & 0.150 & 0.153 & 0.139 & 0.167 & 0.147 \\
     	  & 7 &0.192 & 0.216 & 0.176 & 0.201 & 0.215 & 0.158 & 0.177 & 0.152 & 0.157 & 0.164 \\
     	 & 8 & 0.199 & 0.212 & 0.193 & 0.196 & 0.200 & 0.145 & 0.153 & 0.151 & 0.158 & 0.157 \\
     	  & 9 &0.191 & 0.201 & 0.197 & 0.185 & 0.227 & 0.160 & 0.150 & 0.153 & 0.152 & 0.158 \\
     	  & 10 &0.194 & 0.202 & 0.165 & 0.183 & 0.256 & 0.154 & 0.167 & 0.149 & 0.162 & 0.181 \\
 \hline  
     	   \textbf{Tr2}  & 1 &0.193 & 0.221 & 0.185 & 0.165 & 0.236 & 0.155 & 0.159 & 0.172 & 0.150 & 0.159 \\
     	 (-0.618) & 2 &0.224 & 0.209 & 0.157 & 0.195 & 0.215 & 0.151 & 0.158 & 0.144 & 0.162 & 0.163 \\
     	  & 3 &0.201 & 0.191 & 0.235 & 0.198 & 0.176 & 0.171 & 0.168 & 0.176 & 0.169 & 0.163 \\
     	  & 4 &0.272 & 0.220 & 0.159 & 0.161 & 0.188 & 0.163 & 0.162 & 0.146 & 0.145 & 0.153 \\
     	  & 5 &0.153 & 0.184 & 0.220 & 0.228 & 0.215 & 0.147 & 0.163 & 0.178 & 0.178 & 0.173 \\
     	  & 6 &0.202 & 0.215 & 0.204 & 0.202 & 0.177 & 0.159 & 0.177 & 0.167 & 0.159 & 0.147 \\
     	  & 7 &0.237 & 0.196 & 0.171 & 0.224 & 0.172 & 0.176 & 0.167 & 0.150 & 0.181 & 0.149 \\
     	  & 8 &0.210 & 0.286 & 0.155 & 0.158 & 0.190 & 0.159 & 0.158 & 0.146 & 0.137 & 0.154 \\
     	  & 9 &0.220 & 0.183 & 0.182 & 0.167 & 0.248 & 0.145 & 0.136 & 0.137 & 0.136 & 0.164 \\
     	  & 10 &0.219 & 0.245 & 0.177 & 0.183 & 0.176 & 0.153 & 0.171 & 0.151 & 0.158 & 0.155 \\
 \hline  
     	   \textbf{Tr3}  & 1 &0.163 & 0.249 & 0.186 & 0.163 & 0.240 & 0.139 & 0.171 & 0.150 & 0.134 & 0.163 \\
     	 (0.236) & 2 &0.229 & 0.207 & 0.129 & 0.234 & 0.201 & 0.176 & 0.175 & 0.131 & 0.173 & 0.167 \\
     	 & 3 & 0.239 & 0.196 & 0.198 & 0.167 & 0.199 & 0.162 & 0.152 & 0.165 & 0.146 & 0.151 \\
     	  & 4 &0.212 & 0.231 & 0.199 & 0.161 & 0.196 & 0.159 & 0.170 & 0.157 & 0.141 & 0.164 \\
     	  & 5 &0.181 & 0.229 & 0.170 & 0.206 & 0.213 & 0.162 & 0.178 & 0.134 & 0.164 & 0.159 \\
     	  & 6 &0.244 & 0.208 & 0.204 & 0.177 & 0.168 & 0.176 & 0.168 & 0.164 & 0.149 & 0.144 \\
     	  & 7 &0.210 & 0.186 & 0.158 & 0.243 & 0.204 & 0.159 & 0.161 & 0.145 & 0.167 & 0.151 \\
     	  & 8 &0.220 & 0.204 & 0.208 & 0.169 & 0.200 & 0.158 & 0.162 & 0.164 & 0.150 & 0.166 \\
     	  & 9 &0.205 & 0.211 & 0.183 & 0.173 & 0.228 & 0.151 & 0.162 & 0.148 & 0.151 & 0.166 \\
     	  & 10 &0.262 & 0.243 & 0.176 & 0.186 & 0.132 & 0.174 & 0.184 & 0.162 & 0.158 & 0.142 \\
 \hline  
     	   \textbf{Tr4}  & 1 &0.181 & 0.225 & 0.174 & 0.161 & 0.259 & 0.157 & 0.166 & 0.152 & 0.145 & 0.179 \\
     	 (1.618) & 2 &0.226 & 0.168 & 0.209 & 0.187 & 0.210 & 0.173 & 0.155 & 0.163 & 0.157 & 0.161 \\
     	  & 3 &0.213 & 0.205 & 0.197 & 0.200 & 0.185 & 0.148 & 0.168 & 0.157 & 0.150 & 0.155 \\
     	  & 4 &0.252 & 0.170 & 0.196 & 0.193 & 0.189 & 0.164 & 0.147 & 0.157 & 0.151 & 0.150 \\
     	  & 5 &0.189 & 0.231 & 0.183 & 0.211 & 0.186 & 0.159 & 0.179 & 0.152 & 0.169 & 0.153 \\
     	  & 6 &0.206 & 0.199 & 0.169 & 0.223 & 0.204 & 0.163 & 0.157 & 0.144 & 0.156 & 0.161 \\
     	  & 7 &0.227 & 0.202 & 0.164 & 0.225 & 0.181 & 0.162 & 0.168 & 0.160 & 0.176 & 0.166 \\
     	  & 8 &0.215 & 0.213 & 0.184 & 0.196 & 0.191 & 0.161 & 0.165 & 0.156 & 0.152 & 0.160 \\
     	  & 9 &0.189 & 0.179 & 0.197 & 0.146 & 0.289 & 0.158 & 0.139 & 0.166 & 0.146 & 0.176 \\
     	  & 10 &0.221 & 0.222 & 0.161 & 0.223 & 0.173 & 0.164 & 0.175 & 0.154 & 0.169 & 0.150 \\
 \hline  
     	  \textbf{Tr5}  & 1 &0.200 & 0.208 & 0.197 & 0.199 & 0.195 & 0.150 & 0.155 & 0.159 & 0.151 & 0.156 \\
     	(4.236) & 2 &0.206 & 0.193 & 0.183 & 0.190 & 0.229 & 0.171 & 0.165 & 0.168 & 0.163 & 0.175 \\
     	  & 3 &0.173 & 0.252 & 0.179 & 0.210 & 0.185 & 0.149 & 0.166 & 0.149 & 0.160 & 0.146 \\
     	  & 4 &0.255 & 0.163 & 0.206 & 0.200 & 0.176 & 0.154 & 0.137 & 0.147 & 0.156 & 0.144 \\
     	  & 5 &0.209 & 0.181 & 0.224 & 0.168 & 0.218 & 0.164 & 0.147 & 0.174 & 0.148 & 0.154 \\
     	  & 6 &0.229 & 0.197 & 0.186 & 0.220 & 0.167 & 0.183 & 0.160 & 0.151 & 0.162 & 0.143 \\
     	  & 7 &0.210 & 0.193 & 0.215 & 0.192 & 0.190 & 0.170 & 0.156 & 0.164 & 0.156 & 0.152 \\
     	  & 8 &0.198 & 0.182 & 0.189 & 0.170 & 0.262 & 0.157 & 0.148 & 0.166 & 0.154 & 0.162 \\
     	  & 9 &0.220 & 0.201 & 0.184 & 0.192 & 0.203 & 0.166 & 0.168 & 0.159 & 0.158 & 0.165 \\
     	  & 10 &0.184 & 0.191 & 0.188 & 0.195 & 0.242 & 0.163 & 0.173 & 0.155 & 0.160 & 0.174 \\
 \hline  
 \end{tabular}
\end{table}

\subsection{Human subject game experiment protocol\label{app:humanexp}}

The experiment was approved by the Experimental Social Science Laboratory of Zhejiang University. The data of controlled treatment ($a= [-4.236,	-0.618,	0.234,	1.618,	4.236]$-Treatment) is from the experiment carried out from March to April in 2022.  The authors confirms that this experiment was performed in accordance with the approved social experiments guidelines and regulations, which follow the regulation of \textbf{experimental economics} protocol \cite{Behavioral2003}. 

A total number of 60 undergraduate and graduate students of Zhejiang University volunteered to serve as the human subjects of this experiment. These students were openly recruited through a web registration system. 

The 60 human subjects (call also as players) were distributed into 10 populations of equal size $N$ = 6. The six players of each population carried one experimental sessions. During the game process the players sited separately in a classroom laboratory, each of which facing a computer screen. They were not allowed to communicate with each other during the whole experimental session. Written instructions were handed out to each player and the rules of the experiment were also orally explained by an experimental instructor. The rules of the experimental session are as follows:
\begin{enumerate}
\item 
Each player plays the 5-strategy game repeatedly with the same other five players. 
\item 
Each player earns virtual points during the experimental session according to the payoff matrix shown in the written instruction.   
\item 
In each game round, each player competes with one player in the other five players as the opponent. 
\item 
Each player has  to make a choice among the four candidate actions “x1”, “x2”, “x3” , “x4”, “x5” . If this time runs out, the player has to make a choice immediately. After a choice has been made it can not be changed. If a player does not make a choice in the given 2 second, its strategy of this round will be set to be the same as its own strategy of its previous round.  
\end{enumerate}
 
\begin{figure}
\centering
\includegraphics[width=.8\textwidth]{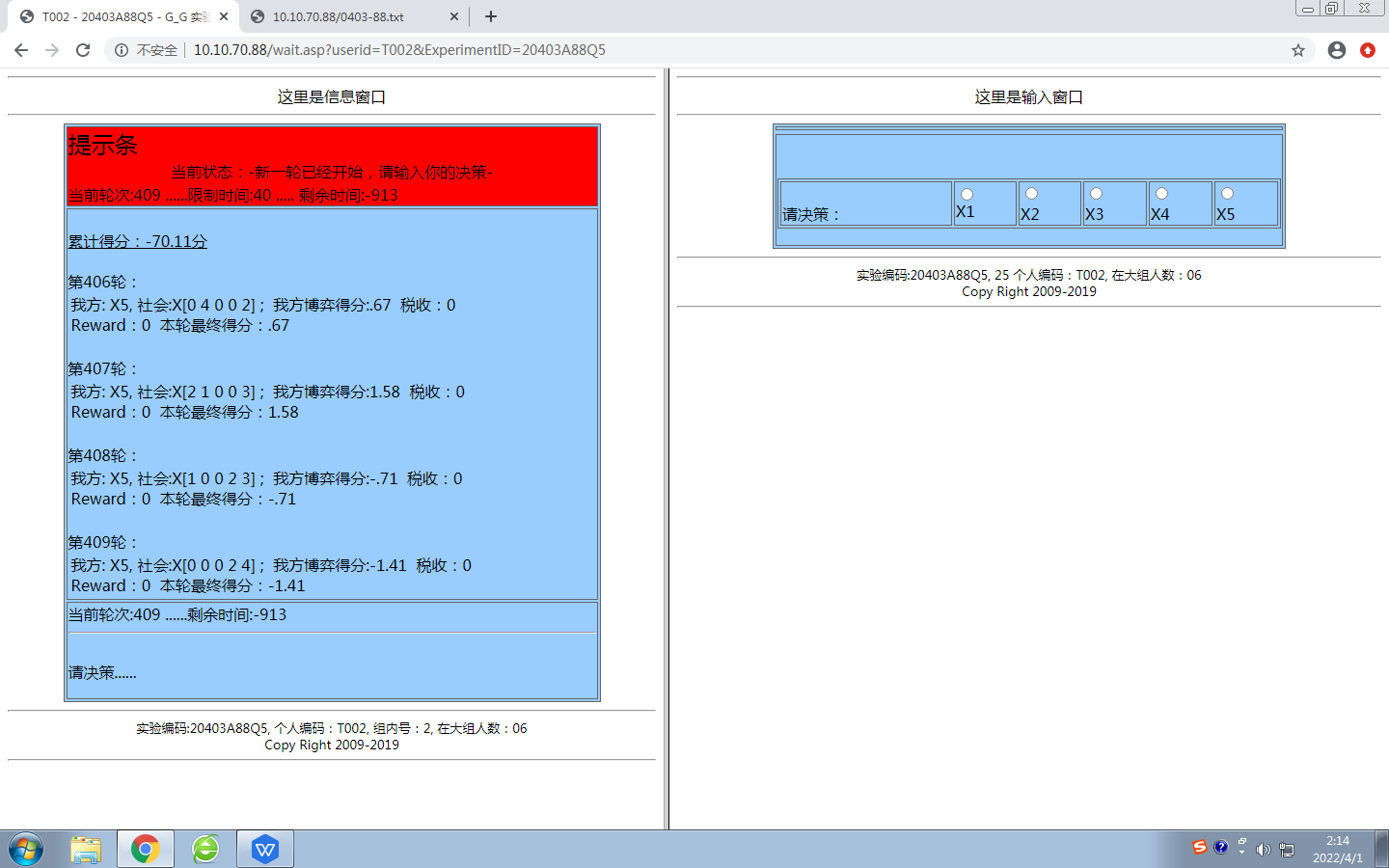}
\caption{\label{fig:expInterface} The screen shot of the user interface in human subject game experiment. }
\end{figure}

During the experimental session, the computer screen of each player will show an information window and a decision window. The window on the left of the computer screen is the information window. The upper panel of this information window shows the current game round, the time limit of making a choice and the time left to make a choice. The color of this upper panel turns to green at the start of each game round.  After a round end, the lower panel of the information window will show the player's own choice, the opponent strategies, and own payoff in this game round are shown in the screen. The opponent strategies is arrange as a list 
 $[P_1,P_2,P_3,P_4,P_5]$ which present the number of players choice  “x1”, “x2”, “x3” , “x4”, “x5” respectively. 

The player's own accumulated payoff is also shown. The players are asked to record by handwriting of their choices of the round on the record sheet in some round for potential checking. Each session last 20 minutes, have more than 600 periods records. For each ($a= [-4.236,	-0.618,	0.234,	1.618,	4.236]$) treatment, we have 10 sessions repeated by various group subjects. Each group was asked to play the 5 sessions without repeated parameter $a$, arranging as $a= [0.234, -0.618,		1.618, -4.236,	4.236]$ in order. So, we have totally 6000 records in the time series for each treatment. 

The window on the right of the computer screen is the decision window. It is activated only after all the players of the group have made their choices. The upper panel of this decision window lists the current game round, while the lower panel lists the five candidate actions (“x1”, “x2”, “x3” , “x4”, “x5”) horizontally from left to right. The player can make a choice by clicking on the corresponding action names.  

The reward for each player is determined by the rank, which is determined by the total number of their earning points in experiment sessions participated. Form the highest to the lowest, each player is payed as 300, 250, 200, 150, 100 and 50 yuan RMB in the sessions participated which cost about 2 hours in total.  
 

\subsection{Additional results}
This sub section includes several statistical results in details, 
for the understanding the main results reported in main text better.

\begin{table} 
\small
 \centering
 \caption{MLE Results By Sessions
 and the statistical ttest By Treatment \label{tab:k1andk2session}}
 \begin{tabular}{cc|rrr|r|rr|rr|}  
 \hline 
($a$) & session&	$c_0$&	$k_\alpha$&	$k_\beta$&$p $~~~&$\bar{k_1}$&$\bar{k_2}$& $p_{k_1}$ &  $p_{k_2}$  \\
 \hline 
     	 \textbf{Tr1} & 1 & -0 & -0.591 & 0.575 & 0 &-0.597 & 0.923 & 0.003 & 0 \\
     	    (-4.236) & 2 & -0 & 0.161 & 1.593 & 0 &&&&\\
     	  & 3 & -0 & -0.401 & 1.168 & 0 &&&&\\
     	  & 4 & -0 & -0.564 & 0.953 &~~~~ 0.003 &&&&\\
     	  & 5 & -0 & -1.676 & 1.199 & 0 &&&&\\
     	  & 6  & -0 & -0.258 & 0.530 & 0.007 &&&&\\
     	  & 7 & 0 & -0.563 & 1.051 & 0.001 &&&&\\
     	  & 8 & -0 & -0.728 & 0.576 & 0.005 &&&&\\
     	  & 9 & -0 & -0.846 & 0.954 & 0 &&&&\\
     	  & 10 & -0 & -0.501 & 0.630 & 0.002 &&&&\\
 \hline 
     	  \textbf{Tr2}  &1 & -0 & 0.479 & 0.048 & 0.001 &0.221 & 0.440 & 0.170 & 0.005 \\
     	 (-0.618) & 2 & -0 & 0.713 & 0.708 & 0 &&&&\\
     	  & 3 & 0 & 0.245 & 0.832 & 0 &&&&\\
     	  & 4 & -0 & 0.125 & 0.114 & 0.505 &&&&\\
     	  & 5  & -0 & -0.222 & 1.229 & 0 &&&&\\
     	  & 6  & 0 & 0.752 & 0.097 & 0.001 &&&&\\
     	  & 7  & 0 & -0.784 & 0.314 & 0.009 &&&&\\
     	  & 8  & -0 & 0.611 & 0.483 & 0.011 &&&&\\
     	  & 9  & 0 & 0.055 & 0.416 & 0.014 &&&&\\
     	  & 10  & 0 & 0.241 & 0.161 & 0 &&&&\\
 \hline 
     	 \textbf{Tr3} &  1 & -0 & -0 & -0.013 & 0.977 &0.293 & 0.134 & 0.004 & 0.093 \\
     	 (0.236) & 2 & 0 & 0.611 & 0.167 & 0.010 &&&&\\
     	  &  3 & 0 & 0.329 & 0.337 & 0.005 &&&&\\
     	  &  4& 0 & 0.070 & 0.401 & 0 &&&&\\
     	  &  5 & 0 & 0.428 & 0.377 & 0.004 &&&&\\
     	  & 6  & -0 & 0.664 & -0.066 & 0.001 &&&&\\
     	  &  7 & 0 & 0.065 & -0.070 & 0.736 &&&&\\
    	  &  8 & 0 & 0.303 & 0.105 & 0.118 &&&&\\
     	  &  9 & 0 & 0.032 & -0.241 & 0.007 &&&&\\
     	  & 10  & -0 & 0.426 & 0.347 & 0.002 &&&&\\
 \hline 
     	 \textbf{Tr4} & 1  & -0 & 1.026 & -0.069 & 0.004 &0.707 & 0.116 & 0 & 0.155 \\
     	 (1.618) &2  & -0 & 0.975 & 0.350 & 0 &&&&\\
     	  & 3  & 0 & 0.735 & 0.319 & 0 &&&&\\
     	  &  4 & 0 & 0.463 & 0.041 & 0.002 &&&&\\
     	  & 5  & -0 & 0.823 & 0.441 & 0.004 &&&&\\
     	  & 6  & -0 & 0.758 & -0.164 & 0 &&&&\\
     	  & 7  & -0 & -0.272 & 0.302 & 0.012 &&&&\\
     	  & 8  & -0 & 1.193 & 0.164 & 0 &&&&\\
     	  & 9  & -0 & 0.765 & -0.255 & 0.013 &&&&\\
     	  & 10  & -0 & 0.605 & 0.028 & 0.001 &&&&\\
 \hline 
     	 \textbf{Tr5} & 1  & -0 & 2.104 & 0.148 & 0 &1.467 & -0.094 & 0 & 0.409 \\
     	 (4.236) & 2 & 0 & 1.704 & 0.479 & 0 &&&&\\
     	  &  3 & -0 & 0.985 & 0.018 & 0.003 &&&&\\
     	  &  4 & -0 & 1.859 & -0.563 & 0 &&&&\\
     	  &  5 & 0 & 0.786 & -0.030 & 0.129 &&&&\\
     	  &  6 & -0 & 1.460 & -0.216 & 0 &&&&\\
     	  &  7 & -0 & 0.434 & 0.251 & 0.206 &&&&\\
     	  &  8 & -0 & 1.606 & -0.145 & 0 &&&&\\
     	  &  9 & 0 & 1.602 & -0.618 & 0 &&&&\\
     	  & 10  & -0 & 2.127 & -0.266 & 0 &&&&\\
 \hline  
 \end{tabular}
\end{table}

\begin{table} 
 
 \centering
 \caption{MLE Results of $c_0$, $k_\alpha$ and $k_\beta$ By Sessions\label{tab:k1__k2sess}}
 \begin{tabular}{cc|rrrr|r||rrrr|r|}  
 \hline 
 $(a)$ & session &	$c_{0a}$&	$c_{0as}$&	$k_a$&	$k_{as}$&	$p_a$&  	$c_{0b}$&	$c_{0bs}$&	$k_b$&	$k_{bs}$&	$p_b$\\
 \hline  

 \textbf{Tr1}&1 &	0.181&	-1&	-0.559&	-1&	0.052&	0.646&	-1&	0.549&	1&	0.03\\
(-4.2362) &2 &	0.501&	-1&	0.249&	-1&	0.705&	-0.176&	-1&	1.6&	1&	0\\
  &3 &	0.368&	-1&	-0.337&	-1&	0.505&	0.438&	-1&	1.15&	1&	0\\
& 4& 0.3&	-1&	-0.511&	-1&	0.262&	0.616&	-1&	0.927&	1&	0.007\\
  &5 &	0.378&	-1&	-1.61&	1&	0.01&	1.831&	-1&	1.124&	-1&	0.072\\
 &6 &	0.167&	-1&	-0.229&	-1&	0.371&	0.282&	-1&	0.518&	1&	0.006\\
 &7 &	0.331&	-1&	-0.505&	-1&	0.286&	0.615&	-1&	1.025&	1&	0.002\\
 &8 &	0.181&	-1&	-0.696&	1&	0.041&	0.795&	-1&	0.543&	-1&	0.09\\
&9  &	0.3&	-1&	-0.793&	-1&	0.069&	0.924&	-1&	0.916&	1&	0.011\\
 &10 &	0.198&	-1&	-0.467&	-1&	0.131&	0.548&	-1&	0.607&	1&	0.015\\
 \hline 
 \textbf{Tr2}  &1 &	0.015&	-1&	0.482&	1&	0&	-0.524&	-1&	0.069&	-1&	0.683\\
(-0.618) &2 &	0.223&	-1&	0.752&	1&	0.032&	-0.779&	-1&	0.741&	1&	0.016\\
 &3 &	0.262&	-1&	0.291&	-1&	0.411&	-0.267&	-1&	0.843&	1&	0\\
&4  &	0.036&	-1&	0.131&	-1&	0.395&	-0.136&	-1&	0.12&	-1&	0.388\\
&5  &	0.387&	-1&	-0.154&	-1&	0.764&	0.243&	-1&	1.219&	1&	0\\
&6  &	0.031&	-1&	0.757&	1&	0&	-0.821&	-1&	0.131&	-1&	0.629\\
&7  &	0.099&	-1&	-0.767&	1&	0.007&	0.857&	-1&	0.278&	-1&	0.377\\
&8  &	0.152&	-1&	0.638&	1&	0.041&	-0.668&	-1&	0.511&	-1&	0.081\\
&9  &	0.131&	-1&	0.078&	-1&	0.7&	-0.06&	-1&	0.419&	1&	0.003\\
&10  &	0.051&	-1&	0.25&	1&	0.006&	-0.263&	-1&	0.172&	-1&	0.062\\
 \hline 
  \textbf{Tr3}&1&	-0.004&	-1&	-0.001&	-1&	0.986&	0&	-1&	-0.013&	-1&	0.823\\
(0.236)&2&	0.053&	-1&	0.62&	1&	0.004&	-0.667&	-1&	0.195&	-1&	0.427\\
&3  &	0.106&	-1&	0.348&	-1&	0.065&	-0.36&	-1&	0.352&	1&	0.032\\
 &4 &	0.126&	-1&	0.092&	-1&	0.595&	-0.076&	-1&	0.405&	1&	0\\
&5  &	0.119&	-1&	0.449&	1&	0.041&	-0.468&	-1&	0.396&	1&	0.048\\
&6  &	-0.021&	-1&	0.66&	1&	0&	-0.725&	-1&	-0.036&	-1&	0.877\\
 &7 &	-0.022&	-1&	0.061&	-1&	0.62&	-0.071&	-1&	-0.067&	-1&	0.545\\
&8  &	0.033&	-1&	0.309&	1&	0.05&	-0.331&	-1&	0.118&	-1&	0.453\\
 &9 &	-0.076&	-1&	0.018&	-1&	0.87&	-0.034&	-1&	-0.239&	1&	0.001\\
 &10 &	0.109&	-1&	0.445&	1&	0.028&	-0.466&	-1&	0.366&	-1&	0.053\\
 \hline 
 \textbf{Tr4}&1&	-0.022&	-1&	1.023&	1&	0.001&	-1.121&	-1&	-0.023&	-1&	0.952\\
(1.618)&2&	0.11&	-1&	0.994&	1&	0.001&	-1.065&	-1&	0.394&	-1&	0.265\\
 &3 &	0.101&	-1&	0.752&	1&	0.002&	-0.803&	-1&	0.353&	-1&	0.198\\
 &4 &	0.013&	-1&	0.465&	1&	0&	-0.505&	-1&	0.062&	-1&	0.711\\
 &5 &	0.139&	-1&	0.848&	1&	0.008&	-0.899&	-1&	0.479&	-1&	0.156\\
 &6 &	-0.051&	-1&	0.749&	1&	0&	-0.828&	-1&	-0.129&	-1&	0.621\\
 &7 &	0.095&	-1&	-0.256&	-1&	0.127&	0.298&	-1&	0.29&	1&	0.044\\
 &8 &	0.052&	-1&	1.202&	1&	0&	-1.304&	-1&	0.218&	-1&	0.59\\
 &9 &	-0.08&	-1&	0.751&	1&	0.006&	-0.835&	-1&	-0.22&	-1&	0.474\\
&10 &	0.009&	-1&	0.606&	1&	0&	-0.661&	-1&	0.055&	-1&	0.797\\
 \hline 
 \textbf{Tr5}&1&	0.047&	-1&	2.112&	1&	0&	-2.298&	-1&	0.243&	-1&	0.734\\
(4.236)&2&	0.151&	-1&	1.73&	1&	0&	-1.861&	-1&	0.556&	-1&	0.352\\
 &3 &	0.006&	-1&	0.986&	1&	0&	-1.076&	-1&	0.062&	-1&	0.863\\
 &4 &	-0.177&	-1&	1.828&	1&	0&	-2.031&	-1&	-0.479&	-1&	0.462\\
 &5 &	-0.01&	-1&	0.785&	1&	0.036&	-0.859&	-1&	0.005&	-1&	0.99\\
&6 &	-0.068&	-1&	1.448&	1&	0&	-1.595&	-1&	-0.15&	-1&	0.769\\
 &7 &	0.079&	-1&	0.448&	-1&	0.13&	-0.474&	-1&	0.27&	-1&	0.333\\
 &8 &	-0.046&	-1&	1.598&	1&	0&	-1.755&	-1&	-0.072&	-1&	0.893\\
 &9 &	-0.195&	-1&	1.568&	1&	0&	-1.75&	-1&	-0.546&	-1&	0.327\\
 &10 &	-0.084&	-1&	2.112&	1&	0&	-2.323&	-1&	-0.17&	-1&	0.814\\

 \hline  
 \end{tabular}
\end{table}

\begin{table} 
 
 \centering
 \caption{Correlation coefficient between experimental $L$ (Change continuously)\label{tab:corr_nTE2}}
 \begin{tabular}{r|rr|rrrrr}  
 \hline 
   Theory $\rho$ & $\sigma^\alpha$ & $\sigma^\beta$ & $M^1$ &  $M^2$  &  $M^3$  &  $M^4$  &  $M^5$  \\  
   \hline
$\sigma^\alpha$ &    1 & 0.0500 & -0.4543 & 0.1307 & 0.8348 & 0.9950 & 0.9950 \\
 $\sigma^\beta$ &   0.0500 & 1 & 0.8670 & 0.9967 & 0.5915 & -0.0498 & -0.0498 \\
 $M^1$ &    -0.4543 & \colorbox[rgb]{1,0.9,0.9}{0.8670} & 1 & 0.8238 & 0.1111 & -0.5408 & -0.5408 \\
 $M^2$ &    0.1307 & \colorbox[rgb]{1,0.9,0.9}{0.9967} & 0.8238 & 1 & 0.6548 & 0.0313 & 0.0313 \\
 $M^3$ &    \colorbox[rgb]{1,0.9,0.9}{0.8348} & \colorbox[rgb]{1,0.9,0.9}{0.5915} & 0.1111 & 0.6548 & 1 & 0.7759 & 0.7759 \\
 $M^4$ &    \colorbox[rgb]{1,0.9,0.9}{0.9950} & -0.0498 & -0.5408 & 0.0313 & 0.7759 & 1 & 1 \\
  $M^5$ &   \colorbox[rgb]{1,0.9,0.9}{0.9950} & -0.0498 & -0.5408 & 0.0313 & 0.7759 & 1 & 1 \\ 
   \hline 	 
   \hline  
Experiment  $\rho$  & $\sigma^\alpha$ & $\sigma^\beta$ & $L^1$ &  $L^2$  &  $L^3$  &  $L^4$  &  $L^5$ \\
   \hline


$\sigma^\alpha$ &       	 1 & 0.050 & -0.470 & 0.426 & 0.866 & 0.970 & 0.996 \\
$\sigma^\beta$ &       	 0.050 & 1 & 0.852 & 0.863 & 0.471 & 0.222 & -0.021 \\
  $L^1$ &     	 -0.470 & 0.852 & 1 & 0.554 & -0.036 & -0.302 & -0.527 \\
 $L^2$ &     	 0.426 & 0.863 & 0.554 & 1 & 0.681 & 0.595 & 0.373 \\
 $L^3$ &      	 0.866 & 0.471 & -0.036 & 0.681 & 1 & 0.922 & 0.824 \\
  $L^4$ &     	 0.970 & 0.222 & -0.302 & 0.595 & 0.922 & 1 & 0.952 \\
 $L^5$ &       	 0.996 & -0.021 & -0.527 & 0.373 & 0.824 & 0.952 & 1 \\
   \hline
Experiment  $p$& $\sigma^\alpha$ & $\sigma^\beta$ & $L^1$ &  $L^2$  &  $L^3$  &  $L^4$  &  $L^5$\\
   \hline   
$\sigma^\alpha$ &     	 0 & 0.891 & 0.171 & 0.219 & 0.001 & 0 & 0 \\
$\sigma^\beta$ &     	 0.891 & 0 & 0.002 & 0.001 & 0.169 & 0.537 & 0.954 \\
 $L^1$ &      	 0.171 & 0.002 & 0 & 0.096 & 0.921 & 0.396 & 0.118 \\
 $L^2$ &     	 0.219 & 0.001 & 0.096 & 0 & 0.030 & 0.070 & 0.289 \\
   $L^3$ &    	 0.001 & 0.169 & 0.921 & 0.030 & 0 & 0 & 0.003 \\
  $L^4$ &     	 0 & 0.537 & 0.396 & 0.070 & 0 & 0 & 0 \\
 $L^5$ &       	 0 & 0.954 & 0.118 & 0.289 & 0.003 & 0 & 0 \\
   \hline
 \end{tabular}
\end{table}

\begin{table} 
 \centering
 \caption{(correlation coefficient of observed eigencycle between 10 2-dimensional subspace \label{tab:corr_e10xe10_full}}
 \footnotesize
\begin{tabular}{r|rrrrrrrrrr} 
\hline 
 & $L^{12}$ & $L^{13}$ & $L^{14}$ & $L^{15}$ & $L^{23}$ & $L^{24}$ & $L^{25}$ & $L^{34}$ & $L^{35}$ & $L^{45}$ \\
 \hline 
$\rho$ \\
 \hline  
 $L^{12}$ &  1& 0.1977& -0.2106& -0.9101& 0.938& 0.3283& -0.2357& 0.8823& 0.3181& 0.8826\\
$L^{13}$ &  0.1977& 1& -0.5832& -0.3445& 0.1704& 0.5599& -0.5855& 0.2655& 0.7611& 0.23\\
$L^{14}$ &  -0.2106& -0.5832& 1& -0.0211& -0.1448& -0.621& 0.5656& -0.1799& -0.4853& 0.024\\
$L^{15}$ &  -0.9101& -0.3445& -0.0211& 1& -0.8719& -0.2554& 0.207& -0.8455& -0.3974& -0.926\\
$L^{23}$ &  0.938& 0.1704& -0.1448& -0.8719& 1& 0.139& -0.3082& 0.9427& 0.2969& 0.8826\\
$L^{24}$ &  0.3283& 0.5599& -0.621& -0.2554& 0.139& 1& -0.7145& 0.1192& 0.5719& 0.2802\\
$L^{25}$ &  -0.2357& -0.5855& 0.5656& 0.207& -0.3082& -0.7145& 1& -0.2813& -0.6138& -0.3258\\
$L^{34}$ &  0.8823& 0.2655& -0.1799& -0.8455& 0.9427& 0.1192& -0.2813& 1& 0.1586& 0.9102\\
$L^{35}$ &  0.3181& 0.7611& -0.4853& -0.3974& 0.2969& 0.5719& -0.6138& 0.1586& 1& 0.1824\\
$L^{45}$ &  0.8826& 0.23& 0.024& -0.926& 0.8826& 0.2802& -0.3258& 0.9102& 0.1824& 1\\
 \hline 
$p$ \\
 \hline 
$L^{12}$ &     0 & 0.1687 & 0.1421 & 0 & 0 & 0.0199 & 0.0994 & 0 & 0.0244 & 0 \\
$L^{13}$ &     0.1687 & 0 & 0 & 0.0143 & 0.2367 & 0 & 0 & 0.0624 & 0 & 0.1082 \\
$L^{14}$ &     0.1421 & 0 & 0 & 0.8842 & 0.3156 & 0 & 0 & 0.2113 & 0.0004 & 0.8687 \\
$L^{15}$ &     0 & 0.0143 & 0.8842 & 0 & 0 & 0.0734 & 0.1491 & 0 & 0.0043 & 0 \\
$L^{23}$ &     0 & 0.2367 & 0.3156 & 0 & 0 & 0.3356 & 0.0294 & 0 & 0.0363 & 0 \\
$L^{24}$ &     0.0199 & 0 & 0 & 0.0734 & 0.3356 & 0 & 0 & 0.4098 & 0 & 0.0487 \\
$L^{25}$ &     0.0994 & 0 & 0 & 0.1491 & 0.0294 & 0 & 0 & 0.0478 & 0 & 0.0210 \\
$L^{34}$ &     0 & 0.0624 & 0.2113 & 0 & 0 & 0.4098 & 0.0478 & 0 & 0.2712 & 0 \\
$L^{35}$ &     0.0244 & 0 & 0.0004 & 0.0043 & 0.0363 & 0 & 0 & 0.2712 & 0 & 0.2050 \\
$L^{45}$ &    0 & 0.1082 & 0.8687 & 0 & 0 & 0.0487 & 0.0210 & 0 & 0.2050 & 0 \\
\hline
  \end{tabular}
\end{table}

\end{document}